\documentclass[twocolumn]{aastex7}

\accepted{December 3, 2025} 

\usepackage{subcaption}
\usepackage{textcomp, gensymb}

\begin{document}

\title{An Analysis of AGN Feedback in the Compact Galaxy Group Stephan's Quintet} 

\author[0009-0005-3001-9989]{Maura Kathleen Shea}
\affiliation{Department of Physics and Astronomy, Georgia State University, 25 Park Place, Atlanta, GA 30303, USA}\email[show]{mshea3@gsu.edu}  

\author[0000-0002-6465-3639]{D. Michael Crenshaw}
\affiliation{Department of Physics and Astronomy, Georgia State University, 25 Park Place, Atlanta, GA 30303, USA}\email{dcrenshaw@gsu.edu}  

\author[0000-0002-3365-8875]{Travis C. Fischer}
\affiliation{AURA for ESA, Space Telescope Science Institute, 3700 San Martin Drive, Baltimore, MD 21218, USA}\email{tfischer@stsci.edu} 

\author[0000-0002-4917-7873]{Mitchell Revalski}
\affiliation{Space Telescope Science Institute, 3700 San Martin Drive, Baltimore, MD 21218, USA}\email{mrevalski@stsci.edu} 

\author[0000-0001-7238-7062]{Julia Falcone}
\affiliation{Department of Physics and Astronomy, Georgia State University, 25 Park Place, Atlanta, GA 30303, USA}\email{jfalcone2@gsu.edu} 

\author[0000-0001-8658-2723]{Beena Meena}
\affiliation{Space Telescope Science Institute, 3700 San Martin Drive, Baltimore, MD 21218, USA}\email{bmeena@stsci.edu} 

\author[0000-0003-3401-3590]{Zo Chapman}
\affiliation{College of Computer Science, Georgia Institute of Technology, 266 Ferst Drive, Atlanta, GA 30332, USA}\email{zchapman9@gatech.edu} 

\author[0000-0002-2713-8857]{Jacob Tutterow}
\affiliation{Kapteyn Astronomical Institute, University of Groningen, P.O. Box 800, 9700 AV Groningen, the Netherlands}\email{jtutterow1@gsu.edu} 

\author[0009-0005-2145-4647]{Madeline Davis}
\affiliation{Department of Physics and Astronomy, Georgia State University, 25 Park Place, Atlanta, GA 30303, USA}\email{mdavis299@gsu.edu} 

\author[0009-0009-4283-3311]{Kesha Patel}
\affiliation{Department of Physics and Astronomy, Tufts University, 574 Boston Avenue,
Medford, MA 02155, USA}\email{kesha.patel@emory.edu}

\begin{abstract}
Compact galaxy groups are ideal laboratories for studying the effects of interactions between AGN and multiple nearby galaxies. 
Recent JWST observations of the nearby compact group Stephan’s Quintet highlight tidal flows between the interacting galaxies as well as outflows from the active galaxy NGC~7319. 
To study the kinematics on a large scale throughout the group, we obtained spatially-resolved long-slit spectra of Stephan’s Quintet at multiple slit positions with Apache Point Observatory's Kitt Peak Ohio State Multi-Object Spectrograph. 
We fit multiple Gaussians to the H$\alpha$ $\lambda$6563 \AA\ and [N~II] $\lambda\lambda$6548, 6583 \AA\ emission lines to isolate the different kinematic components. We used the kinematics to develop the first biconical outflow model of the narrow-line region of NGC~7319. 
Using a combination of galactic rotation models, biconical outflow models, and kinematic maps of the ionized gas, we disentangled the outflows, rotation, and tidal flows in the group. 
We found outflow radial velocities up to 550 km~s$^{-1}$ peaking at 2.6 kpc from the central supermassive black hole, and a transition from AGN-powered outflows to gravitationally-powered tidal flows at a projected distance between 2.4~--~6.3~kpc. 
We performed a line ratio analysis and determined the gas shows Seyfert-like ionization out to 6.3 kpc (projected), which supports our finding that gas outside this radius is predominantly powered by tidal flows. 
Our separation of kinematic components in Stephan’s Quintet will enable future studies of the physical conditions and dynamical forces in the ionized gas to better quantify the feeding and feedback processes of AGN in compact groups. 
\end{abstract}


\section{Introduction} \label{sec:intro}
Nearly all massive galaxies are expected to contain supermassive black holes (SMBHs) at their centers \citep{kormendy_coevolution_2013}. Of these black holes, approximately 5$-$10\% are active at the present epoch \citep{peterson_introduction_1997}, emitting powerful radiation and mass outflows of gas \citep{fabian_observational_2012}. These active galactic nuclei (AGN) are powered by the infall and the subsequent heating of matter in an accretion disk around the central SMBH. Some of the infalling mass is pushed outward by radiation or magnetic fields \citep{fabian_observational_2012}, resulting in mass outflows. In moderate to high luminosity (L$_{bol}$~$\geq$~10$^{43}$~erg~s$^{-1}$) AGN, the outflowing gas is primarily ionized by radiation released from the accretion disk \citep{ma_spatially_2021, kraemer_iue_1994, kraemer_probing_2008, kraemer_physical_2009}, although other ionizing sources such as shocks or stellar can be present \citep{zhang_modelling_2013,dagostino_separating_2019}, allowing us to determine the gas kinematics and physical conditions of these AGN outflows via emission line spectroscopy.

AGN outflows typically take the form of a biconical structure, with the peak of the bicone located at the central AGN \citep{antonucci_spectropolarimetry_1985, pogge_extended_1988, pedlar_radio_1993, schmitt_anisotropic_1994, nelson_space_2000, fischer_determining_2013}.  The bicones are formed by the torus, a ring of dusty gas outside of the accretion disk that obscures our view of the central AGN in Seyfert~2s (whereas Seyfert~1s have an unobscured view along our line of sight).
The narrow-line regions (NLRs) in Seyferts typically exhibit well-defined edges and a broad range of projected opening angles, ranging from $\sim$30\degree$-$100\degree\ \citep{schmitt_hubble_2003, fischer_determining_2013}. Understanding the biconical structure and kinematics of an AGN NLR can help us understand its geometry, interaction with the host galaxy, fueling flow, and impact on its environment.

AGN feedback can greatly affect the host galaxy by ionizing and removing gas directly from cold gas reservoirs and by depositing energy and momentum from the resulting outflows into the interstellar medium. These processes can result in the expulsion and heating of large quantities of gas. The effects of these outflows can extend far beyond the NLR: outflows play an essential role in the development of large-scale structure in the early Universe \citep{scannapieco_quasar_2004, di_matteo_energy_2005}, in self-regulation of SMBH and galactic bulge growth \citep{hopkins_black_2005, hopkins_quasar_2010}, in galaxy evolution \citep{venturi_magnum_2021}, and in chemical enrichment of the intergalactic medium \citep{khalatyan_is_2008} as well as the intragroup medium (IGrM) in galaxy groups or clusters. Furthermore, it is well known that AGN outflows can limit and even potentially quench star formation in the host galaxy in a form of ``negative
feedback" \citep{fabian_cooling_1994, fischer_gemini_2017, fischer_hubble_2018, revalski_quantifying_2021}.
However, AGN outflows can also trigger star formation through the compression
of the interstellar gas, causing ``positive feedback" \citep{silk_global_2009, silk_unleashing_2013, van_dokkum_substantial_2010, shin_positive_2019}. While these processes are under intense study in clusters of galaxies \citep{masterson_evidence_2023, saxena_widespread_2024, spasic_victoria_2024}, little has been documented about AGN interactions with the IGrM in compact galaxy groups.

\subsection{Compact Galaxy Groups}

Compact galaxy groups are composed of a few gravitationally bound galaxies in close proximity to one another, forming the densest galaxy systems. They are defined by the criteria established by \cite{hickson_systematic_1982} in his catalog of Hickson Compact Groups (HCGs): 

\begin{enumerate}
    \item The group must contain four or more galaxies whose magnitudes differ by less than 3.0 magnitudes.
     \item The group must have a mean surface brightness brighter than 26.0 mag per arcsec$^2$ in the \textit{g} band.
    \item There cannot be an external galaxy within three radii of the smallest circle that contains the geometric centers of the above galaxies.
\end{enumerate} 
The last criterion was used to ensure the compact group is sufficiently isolated from neighboring galaxies, so that the cores of rich galaxy clusters were not mistakenly included in the catalog.

The majority of compact groups contain galaxies exhibiting morphologic or kinematic peculiarities, as well as high amounts of intergalactic gas and frequent galaxy interactions and mergers \citep{hickson_compact_1997, bitsakis_mid-ir_2010}. On average, 43\% of galaxies in HCGs show signs of past interactions or mergers, and 32\% of HCGs contain at least three interacting galaxies \citep{hickson_compact_1997}. The galaxy interactions can both remove gas from the outer regions of a galaxy, and allow for gas to flow into the galaxy towards the nucleus. Due to these interactions, AGN activity and star formation rates are substantially increased within compact groups \citep{bahar_srgerosita_2024, de_rosa_multiple_2015, plauchu-frayn_star_2012}, with up to 42\% of galaxies in compact groups hosting an AGN \citep{martinez_agn_2010, sohn_activity_2013}. These traits make low-redshift compact groups an excellent environment for the study of extreme galaxy interactions, which are thought to have a significant role in driving galaxy evolution at high redshift \citep{rodriguez-baras_study_2014}, as well as AGN activity in these environments.

\subsection{Stephan's Quintet and NGC~7319}

\begin{figure}
    \centering
    \includegraphics[width=\linewidth]{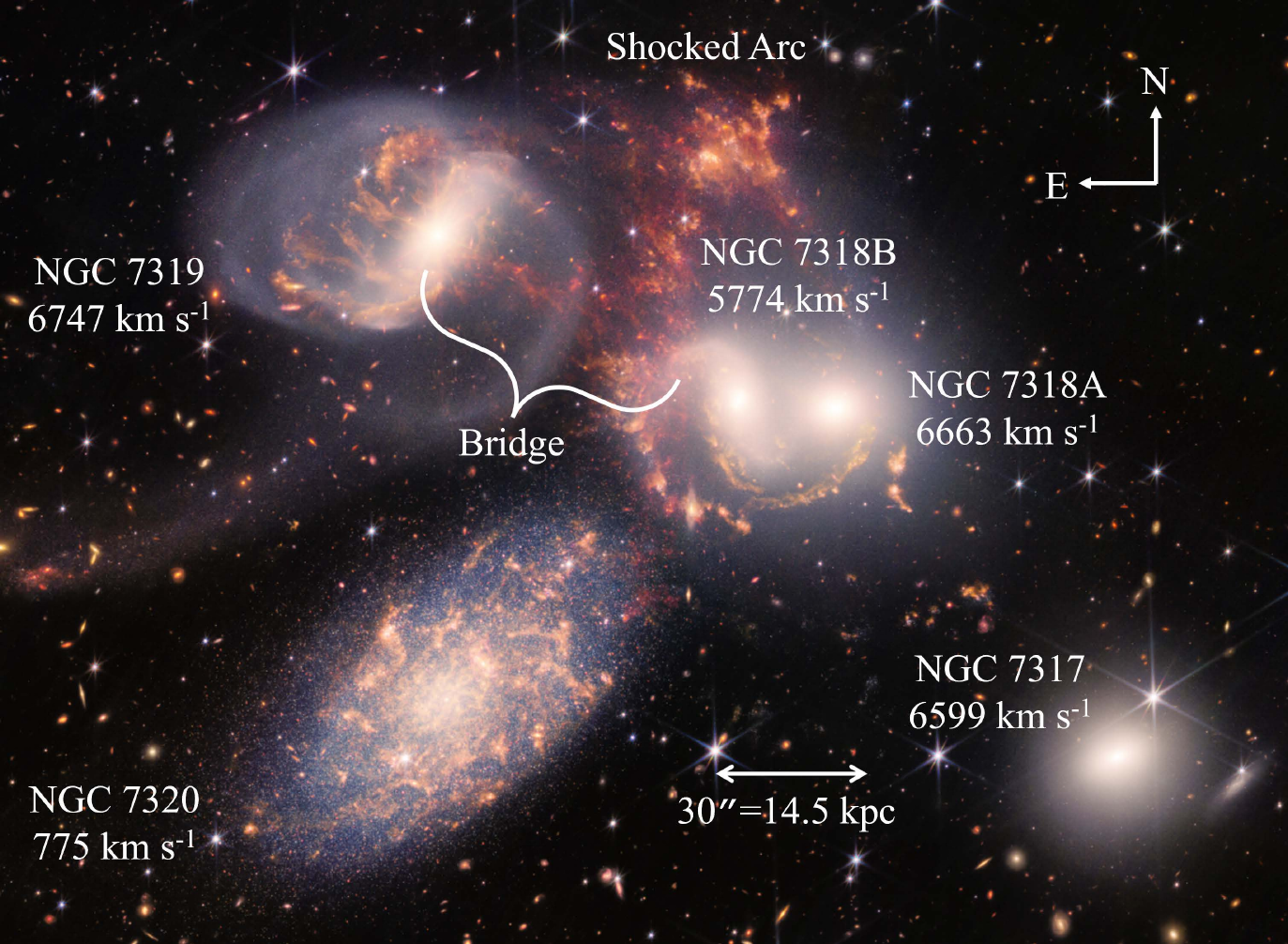}
    \caption[]{Stephan's Quintet, as imaged by the James Webb Space Telescope (\textit{Image credit: NASA, ESA, CSA, and STScI}. Created with data from \dataset[JWST ERO 2732]{https://www.stsci.edu/jwst/science-execution/program-information?id=2732} (PI: Pontoppidan) with NIRCam F090W, F150W, F200W, F277W, F365W, F444W filters and MIRI F770W, F1000W filters).  Cluster members and systemic redshift velocities provided by the NASA/IPAC Extragalactic Database (NED) are labeled. The bridge, a key feature seen later in the kinematic plots (Figure~ \ref{fig:2D_kinematic}), is also labeled. The curved, shocked ridge of ionized gas between NGC~7319 and NGC~7318B is also clearly evident. Scale provided in the image is for NGC~7319.}
    \label{fig:SQ}
\end{figure}


Stephan's Quintet (HCG 92), shown in Figure~\ref{fig:SQ}, is an ideal example of a compact galaxy group as it is nearby (z~=~0.022), bright, and undergoing strong dynamical interactions \citep{rodriguez-baras_study_2014}. Discovered in 1877 by \'Edouard Stephan, it was the first compact group observed \citep{stephan_nebulae_1877, hickson_compact_1997}. Of the five galaxies in the group, three show strong tidal distortions due to gravitational interactions with each other \citep{hickson_compact_1997}, some of which triggered periods of massive star formation \citep{fedotov_star_2011}. Stephan’s Quintet is also home to an AGN at the center of NGC~7319, which possesses large-scale outflows (4~kpc radially) and is expected to have been involved in the majority of past interactions within the group \citep{aoki_high_1996}.

\begin{table}[]
    \centering
    \begin{tabular}{|l|l|l|l|} \hline
        \multicolumn{1}{|c}{Galaxy} & 
        \multicolumn{1}{|c}{Redshift} & 
        \multicolumn{1}{|c}{cz} & 
        \multicolumn{1}{|c|}{Distance} \\

        \multicolumn{1}{|c}{} & 
        \multicolumn{1}{|c}{} & 
        \multicolumn{1}{|c}{[km s$^{-1}$]} & 
        \multicolumn{1}{|c|}{[Mpc]} \\

        \hline
        \hline
        NGC 7317 & 0.022012$\pm$8.70e-5  & 6599$\pm$26 & 92.51$\pm$6.51  \\ 
        NGC 7318A & 0.022225$\pm$7.00e-5 & 6663$\pm$21 & 93.45$\pm$6.56  \\ 
        NGC 7318B & 0.019260$\pm$8.00e-5 & 5774$\pm$24 & 80.34$\pm$5.66  \\ 
        NGC 7319 & 0.022507$\pm$1.20e-5 & 6747$\pm$4 & 	94.70$\pm$6.64  \\ 
        NGC 7320 & 0.002585$\pm$2.00e-6 & \hspace{4pt}775$\pm$1 & \hspace{4pt}6.61$\pm$0.57  \\ 
        \hline
    \end{tabular}
    \caption{Redshift, redshift velocity, and distances for each galaxy in Stephan's Quintet, provided by NED. Distances were calculated assuming H$_0 = 67.8$ km s$^{-1}$.}
    \label{table:SQ}
\end{table}

NGC~7319 (see Table~\ref{table:SQ}, $S~=~482$~pc~arcsec$^{-1}$) is most commonly considered a Seyfert 2 galaxy \citep{veron-cetty_catalogue_2006}. Seyfert galaxies harbor nearby (z $\leq$ 0.1) AGN with moderate luminosities (L$\mathrm{_{bol}}$ $\approx 10^{43}-10^{45}$ erg s$^{-1}$; \citealp{peterson_introduction_1997}). Recent observations of NGC~7319 by \cite{yttergren_gas_2021} claim a possible weak broad component in H$\alpha$, suggesting that NGC~7319 may be a Seyfert 1.9. In our data (Section \ref{sec:obs}) we did not detect any broad lines, and therefore continue to treat NGC~7319 as a Seyfert 2 galaxy in the analysis.

NGC~7319, NGC~7318A, and NGC~7317 make up the core of Stephan's Quintet and have low velocity differences relative to each other (see Figure~\ref{fig:SQ} and Table~\ref{table:SQ}). These galaxies show signs of past dynamical interactions with an ``old intruder" galaxy, NGC~7320C ($cz=5985$ km~s$^{-1}$; not shown in Figure~\ref{fig:SQ}), which collided with the group $\sim$$10^8$ years ago, stripping much of the interstellar medium from NGC~7319 \citep{moles_dynamical_1997}. In the current epoch, NGC~7318B is entering the group for the first time at a high relative velocity ($\sim$1000 km s$^{-1}$), colliding with the IGrM of the group and giving rise to a shock zone (the curved structure to its east in Figure~\ref{fig:SQ}), and a tidal bridge between NGC~7319 and 7318B \citep{sulentic_multiwavelength_2001, xu_physical_2003, rodriguez-baras_study_2014, arnaudova_weave_2024}. NGC~7320 is a foreground spiral galaxy \citep{burbidge_further_1961}. By investigating NGC~7319 in relation to the other members of Stephan's Quintet, we begin to address how an AGN affects (and is effected by) not only its own galaxy, but the entire compact group.

\section{Observations} \label{sec:obs}
\begin{table*}[tt]
\centering
\footnotesize

\begin{tabular} 
{|r r r r r r r r r r r|} 

\hline
\multicolumn{1}{|C}{\textnormal{Target}} & 
\multicolumn{1}{C}{\textnormal{Instrument}} & 
\multicolumn{1}{C}{\textnormal{Program}} & 
\multicolumn{1}{C}{\textnormal{Date}} & 
\multicolumn{1}{C}{\textnormal{Grism/}} & 
\multicolumn{1}{C}{\textnormal{Exposure}} & 
\multicolumn{1}{C}{\textnormal{Wavelength}} & 
\multicolumn{1}{C}{\textnormal{Seeing}} & 
\multicolumn{1}{C}{\textnormal{Spatial}} & 
\multicolumn{1}{C}{\textnormal{Position}} & 
\multicolumn{1}{C|}{\textnormal{Spatial}} \\

& 
& 
\multicolumn{1}{C}{\textnormal{ID}} & 
& 
\multicolumn{1}{C}{\textnormal{Filter}} & 
\multicolumn{1}{C}{\textnormal{Time}} & 
\multicolumn{1}{C}{\textnormal{Range}} & 
\multicolumn{1}{C}{\textnormal{FWHM}} & 
\multicolumn{1}{C}{\textnormal{Scale}} & 
\multicolumn{1}{C}{\textnormal{Angle}} & 
\multicolumn{1}{C|}{\textnormal{Offset}} \\

& 
& 
& 
\multicolumn{1}{C}{\textnormal{(UT)}} & 
& 
\multicolumn{1}{C}{\textnormal{(s)}} & 
\multicolumn{1}{C}{\textnormal{(\AA)}} & 
\multicolumn{1}{C}{\textnormal{(\arcsec)}} & 
\multicolumn{1}{C}{\textnormal{(\arcsec/pix)}} & 
\multicolumn{1}{C}{\textnormal{(deg)}} & 
\multicolumn{1}{C|}{\textnormal{(\arcsec NW)}} \\

\hline
\hline

NGC 7318B & WFC3/UVIS & 11502 & 2009 July 29 & F606W & 5400 & $4795-6985$ & N/A & 0.040 & 0 & N/A \\

NGC 7318B & WFC3/UVIS & 11502  & 2009 July 29 & F814W & 7200 & $7255-8820$ & N/A & 0.040 & 0 & N/A \\

NGC 7318B & WFC3/UVIS & 11502  & 2009 July 30 & F665N & 20800 & $6552-6570$ & N/A & 0.040 & 0 & N/A \\

NGC 7319 & KOSMOS & GS01  & 2022 July 26 & Blue & 2700 & $4150-7050$ & 2.00 & 0.257 & 60 & 0 \\

NGC 7319 & KOSMOS & GS01  & 2022 July 26 & Blue & 2700 & $4150-7050$ & 2.00 & 0.257 & 150 & 0 \\

NGC 7319 & KOSMOS & GS01  & 2022 Oct 25 & Blue & 2700 & $4150-7050$ & 1.38 & 0.257 & 60 & 0 \\

NGC 7319 & KOSMOS & GS01  & 2022 Oct 25 & Blue & 2700 & $4150-7050$ & 1.38 & 0.257 & 60 & -2 \\

NGC 7319 & KOSMOS & GS01  & 2022 Oct 25 & Blue & 2700 & $4150-7050$ & 1.38 & 0.257 & 60 & -4 \\

NGC 7319 & KOSMOS & GS01 & 2022 Oct 25 & Blue & 2700 & $4150-7050$ & 1.38 & 0.257 & 60 & -6 \\

NGC 7319 & KOSMOS & GS01 & 2022 Oct 25 & Blue & 2700 & $4150-7050$ & 1.38 & 0.257 & 60 & -8 \\

NGC 7319 & KOSMOS & GS01  & 2022 Nov 27 & Blue & 2700 & $4150-7050$ & 1.02 & 0.257 & 60 & 0 \\

NGC 7319 & KOSMOS & GS01  & 2022 Nov 27 & Blue & 2700 & $4150-7050$ & 1.02 & 0.257 & 60 & 2 \\

NGC 7319 & KOSMOS & GS01  & 2022 Nov 27 & Blue & 2700 & $4150-7050$ & 1.02 & 0.257 & 60 & 4 \\

NGC 7319 & KOSMOS & GS01  & 2022 Nov 27 & Blue & 2700 & $4150-7050$ & 1.02 & 0.257 & 60 & 6 \\

NGC 7319 & KOSMOS & GS01  & 2022 Nov 27 & Blue & 2700 & $4150-7050$ & 1.02 & 0.257 & 60 & 8 \\

NGC 7319 & KOSMOS & GS01  & 2022 Nov 27 & Blue & 2700 & $4150-7050$ & 1.02 & 0.257 & 60 & 10 \\

NGC 7319 & KOSMOS & GS01  & 2023 July 18 & Red & 3600 & $5600-9400$ & 1.27 & 0.257 & 150 & 0 \\

NGC 7319 & KOSMOS & GS01 & 2023 Dec 07 & Blue & 5400 & $4150-7050$ & 1.40 & 0.257 & 18 & 0 \\

HD 97438 & KOSMOS & GS01  & 2024 April 30 & Red & 270 & $5600-9400$ & 0.86 & 0.257 & 0 & 0 \\

HD 117876 & KOSMOS & GS01  & 2024 April 30 & Red & 90 & $5600-9400$ & 0.86 & 0.257 & 0 & 0 \\

HD 125560 & KOSMOS & GS01  & 2024 April 30 & Red & 3 & $5600-9400$ & 0.86 & 0.257 & 0 & 0 \\

HD 97438 & KOSMOS & GS01  & 2024 June 13 & Blue & 270 & $4150-7050$ & 1.10 & 0.257 & 0 & 0 \\

HD 117876 & KOSMOS & GS01  & 2024 June 13 & Blue & 12 & $4150-7050$ & 1.10 & 0.257 & 0 & 0 \\

HD 125560 & KOSMOS & GS01  & 2024 June 13 & Blue & 6 & $4150-7050$ & 1.10 & 0.257 & 0 & 0 \\

\hline

\end{tabular}

\caption{HST and KOSMOS observations of Stephan's Quintet and template stars.  The columns list (1) primary target, (2) instrument, (3) Program ID, (4) observation date, (5) grating (for spectra) or filter (for imaging), (6) exposure time for each observation, (7) wavelength range (of the spectra) or bandpass (of the filter), (8) the seeing, as determined by the FWHM of a bright guide star, (9) spatial scale of the spectra, (10) position angle of the slits or angle of the images, and (11) the spatial offsets of the slits or images from the nucleus. The values for columns (7) and (9) for the HST data were obtained in their corresponding instrument handbooks \citep{dressel_wide_2012}.}
\label{table:obs}
\end{table*}

\begin{figure*}
    \centering
    \includegraphics[width=\textwidth]{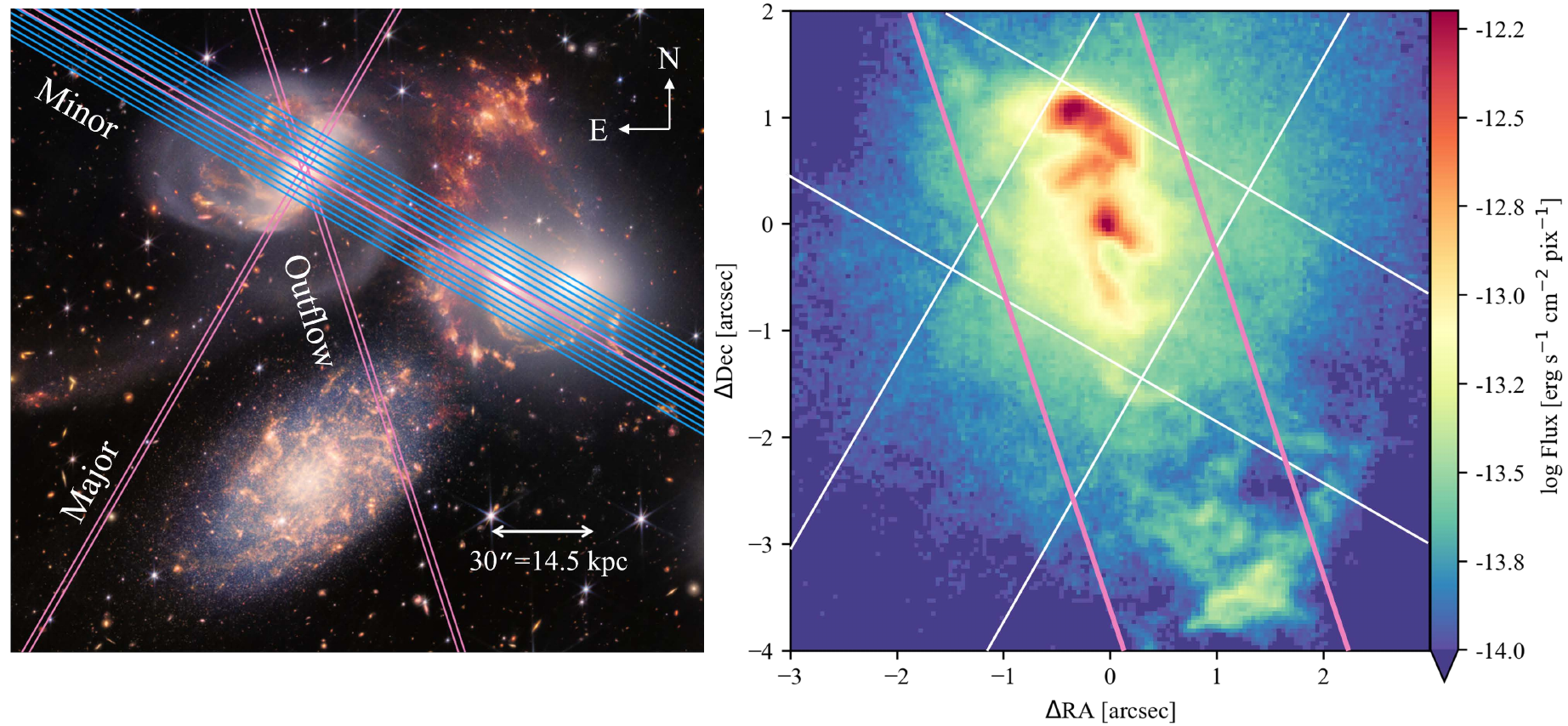}
    \caption{\textit{Left:} Stephan's Quintet with KOSMOS slits overlaid to scale (underlying image same as Figure \ref{fig:SQ}). Pink slits are observations taken along the major (slit PA 150\arcdeg) and minor (PA 60\arcdeg) axes of NGC~7319, and along the outflowing axis of the NLR (PA 18\arcdeg). Blue slits are observations along the minor axis but offset from the nucleus in increments of 2\arcsec. In these observations, the slits are adjacent but not overlapping. We obtained four observations offset to the southeast of the nucleus and five to the northwest. \textit{Right:} A continuum-subtracted H$\alpha$ image from HST WFC3 of the nuclear region of NGC~7319, with the nucleus-centered slits overlaid. The outflowing axis is still highlighted in pink, while the major and minor axes are in white.}
    \label{fig:slits}
\end{figure*}

\subsection{Apache Point Observatory}

 We obtained observations of Stephan’s Quintet using the Kitt Peak Ohio State Multi-Object Spectrograph (KOSMOS) on the Astrophysical Research Consortium (ARC) 3.5-meter telescope at Apache Point Observatory (APO). We used the KOSMOS 6\arcmin\ $\times$ 2\arcsec\ slits, which are ideal for measuring the ionized gas kinematics on large scales as they can span the entire compact group while spatially resolving the emission lines. The position angles of the slits were centered on NGC~7319 and chosen to cover the major and minor axes of NGC~7319's galactic disk -- position angles (PAs) of 150\arcdeg\ and 60\arcdeg, respectively. The major axis PA was determined by the orientation of the disk, while the minor axis is orthogonal. Our measurement is consistent with previous measurements of the major axis PA by \cite{jarrett_2mass_2003}. We later observed the outflow axis of the NLR (PA = 18\arcdeg), which we determined from an HST H$\alpha$ narrow-band image. Additionally, we obtained observations with slit positions offset from the  minor axis, providing coverage of the AGN host galaxy, companion galaxies, and on scales of the IGrM. All slit positions relative to the central galaxies in Stephan's Quintet are shown in Figure~\ref{fig:slits}, and observations are listed in Table~\ref{table:obs}.
 
For KOSMOS, the wavelength coverage is determined by a combination of the grism and slit placement. We used the blue high slit (wavelength range of $\lambda$ = 4150 -- 7050 \AA) to capture the 
H$\beta$ $\lambda$4861 \AA, 
[O~III] $\lambda\lambda$4959, 5007 \AA, 
[N~II] $\lambda\lambda$6548, 6583 \AA, 
H$\alpha$ $\lambda$6563 \AA, 
[O~I]$\lambda\lambda$6300, 6363 \AA, and  
[S~II] $\lambda\lambda$6716, 6731 \AA\ 
emission lines and the 
Mg~I~b $\lambda\lambda\lambda$5167.3, 5172.6, 5183.6 \AA\ 
absorption line. 
We used the red center slit ($\lambda$ = 5600 -- 9400 \AA) to capture the 
Ca~II triplet $\lambda\lambda\lambda$8498.0, 8542.0, 8662.1~\AA\
and 
Na~I absorption lines $\lambda\lambda$5889.9, 5895.9 \AA.
The KOSMOS spectra have a spectral dispersion of 0.71 \AA\ pix$^{-1}$ and a spectral resolution of 5.1 \AA\ (full-width at half-maximum; FWHM) for a fully-illuminated 2\farcs0-wide slit.
 
Using standard IRAF routines involving bias subtraction and flat-field correction, we reduced the data into calibrated spectral images \citep{tody_iraf_1986, hanisch_astronomical_1993}. We observed a standard star selected from the \cite{oke_faint_1990} catalog for flux calibrations each night. In between each science exposure that required the telescope to slew, we took arc lamp images to use for wavelength calibration.  We performed additional calibrations using IDL to correct the tilt of the spectrum in the cross-dispersion direction \citep{gnilka_gemini_2020}.

Truss flats (i.e. flats taken with lamps on the truss of the telescope, external to the instrument) taken with the bright quartz (continuum) lamp include a bright line where a 0$^{th}$-order undispersed spectrum passes through the grism. Due to the particular redshift of NGC~7319, this bright feature aligns with the blue wing of the [O~III] $\lambda$5007 \AA\ emission line. To correct the bright feature in the master flat, we averaged 20 columns on either side of the bright line, then created a linear gradient between these two values. This process was done for each spatial row to create a smooth interpolation. 

\subsection{Hubble Space Telescope}

We used H$\alpha$ images taken by the Wide Field Camera 3 (WFC3) on the Hubble Space Telescope (HST) to assess the structure and direction of the ionized gas outflows.  We retrieved the calibrated data from the Mikulski Archive at the Space Telescope Science Institute (MAST). The images used the F606W, F814W, and F665N filters and were obtained by \dataset[HST PID 11502]{https://archive.stsci.edu/proposal_search.php?mission=hst&id=11502} (PI: Noll). The observations were taken during the same observing session, so no significant spatial misalignment is expected. We began by converting the data from electrons~s$^{-1}$ to erg~cm$^{-2}$~s$^{-1}$ by multiplying the data counts for each image by its filter's inverse sensitivity and its filter bandwidth. To create the continuum-subtracted image, we scaled all fluxes to the width of the narrow-band filter, then took the weighted average of the F606W and the F814W images. We then subtracted this from the F665N data containing H$\alpha$ emission. The resulting image exhibiting the ionized gas is shown in the right panel of Figure~\ref{fig:slits}. The nucleus was determined by the peak in continuum flux, and is positioned at (0,0).

\section{Methods} \label{sec:methods}
 The two-dimensional spectra from KOSMOS cover not only the nucleus of NGC~7319, but also the IGrM and other galaxies in the cluster. From detailed measurements of the spatially-resolved emission and absorption lines, we determined how the kinematics and physical conditions vary across Stephan's Quintet. 

\subsection{Bayesian Evidence Analysis Tool (BEAT)} \label{sec:BEAT}

\begin{figure*}
    \centering
    \includegraphics[width=0.9\textwidth]{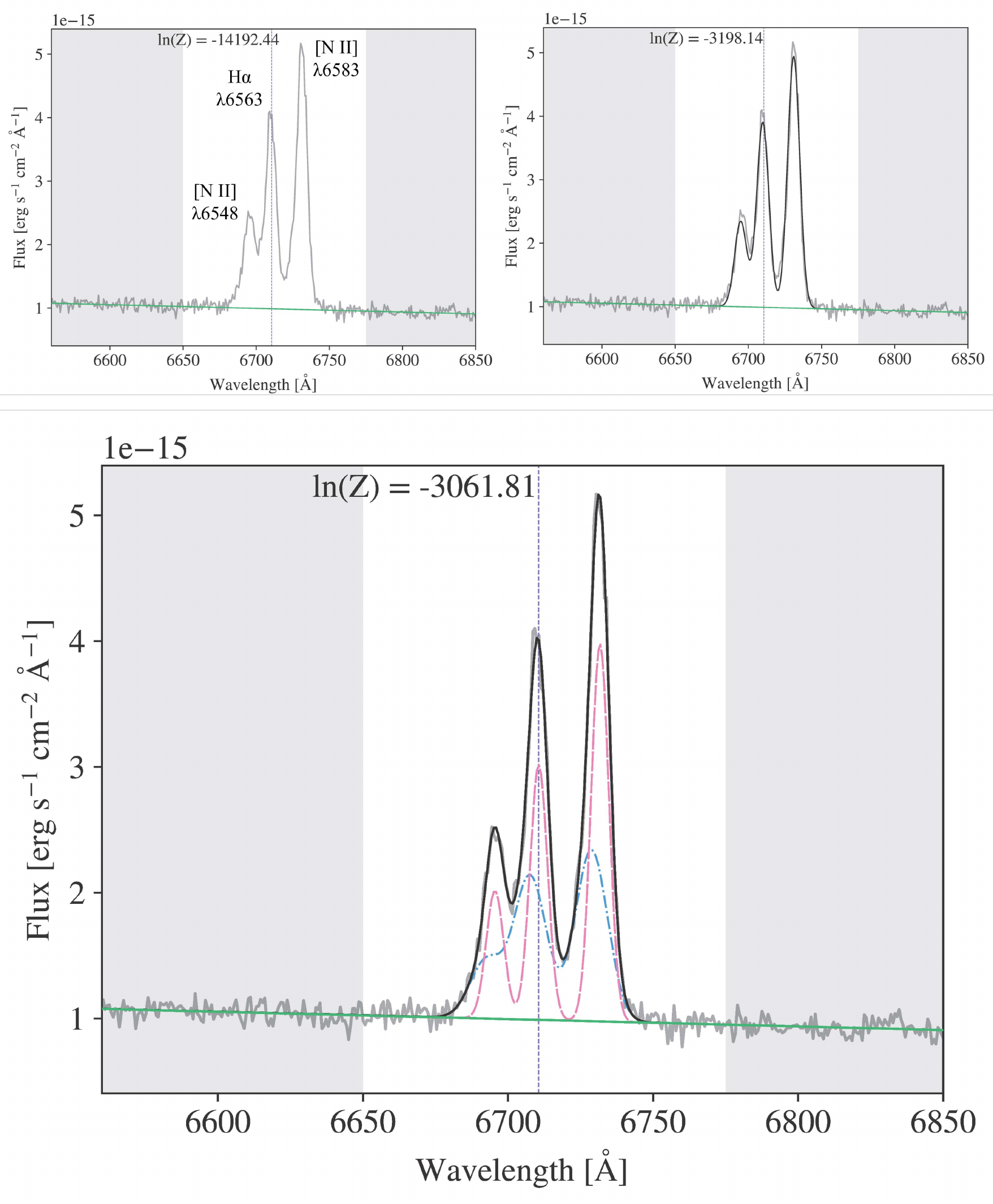}
    \caption{Examples of BEAT fitting the H$\alpha$ $\lambda$6563 \AA\ $+$ [N~II] $\lambda\lambda$6548, 6583 \AA\ emission lines near the nucleus of NGC~7319. The grey curve represents the data, while the black curve shows the composite fit. The green line marks the continuum level, and the grey shaded regions are those chosen to calculate the continuum. The vertical purple dashed line is the redshifted rest wavelength of H$\alpha$. \textit{Top Left:} The zero component fit for the data, with key lines labeled. \textit{Top Right:} The one component fit. \textit{Bottom:} The two component fit. The pink and blue curves are the two individual component fits used to create the composite fit. In this case, BEAT finds the two-component fit to be the best model, as indicated by the difference in ln(\textbf{Z}) values. Not shown is the 3-component fit, which had a more negative ln(\textbf{Z}) than the 2-component fit, and was therefore not chosen.}
    \label{fig:beat}
\end{figure*}

The large wavelength range of KOSMOS encompasses both the  H$\beta$ $\lambda$4861 \AA $+$ [O~III] $\lambda \lambda$4959, 5007 \AA\ region and the H$\alpha$ $\lambda$6563 \AA $+$ [N~II] $\lambda \lambda$6548, 6583 \AA\ region. To characterize the ionized gas kinematics in NGC~7319, we chose to focus on the H$\alpha$ $+$ [N~II] region due to its brightness, ability to trace both star formation and AGN activity, and its lower sensitivity to extinction effects compared to the bluer emission lines.
Figure~\ref{fig:beat} shows this set of lines near the nucleus of NGC~7319, where the emission lines are strongest. At each point along the slit, we fit multiple Gaussian profiles to measure how the velocities of the gas components are changing with position. Emission-line profiles are not always perfect Gaussians \citep{heckman_emission-line_1981, veilleux_study_1991}. However, fitting profiles with multiple Gaussians has been accurately used to determine the velocity centroids and widths of multiple kinematic components in the line of sight, as well as fluxes of the corresponding emission lines \citep{fischer_hubble_2018,revalski_quantifying_2018, revalski_quantifying_2021, gnilka_gemini_2020, meena_radiative_2021, meena_investigating_2023, falcone_analysis_2024}.

To fit Gaussians to the emission lines, we used a fitting routine called the Bayesian Evidence Analysis Tool (BEAT;  \citealp{fischer_gemini_2017, falcone_analysis_2024}). BEAT employs the sampling algorithm MultiNest \citep{feroz_importance_2019} to determine the fewest number of significant kinematic components. Each of these components is characterized as a Gaussian with a velocity centroid, width, and height above the underlying continuum. Physically, each Gaussian component represents a distinct filament, or combination of filaments or knots with similar velocities, in the gas. The wide area covered by the KOSMOS slit means the resulting spectrum at each point along the slit is typically a superposition of multiple knots' motions. Therefore, we expect (and find) that more clumpy or filamentary areas, such as those closer to the AGN, will be fit with more components representing the individual knots as compared to areas farther from the nucleus. 

BEAT requires several Bayesian priors in order to run, including the definition of two continuum regions, line position and width limits, and line ratios for those lines that have fixed ratios according to atomic physics. The continuum regions were chosen to avoid any emission or absorption lines on either side of the feature and fit by a first-order polynomial, as shown by the grey shaded regions in Figure~\ref{fig:beat}.
The lower limit of the line widths was set to 2.4 or 3.8 \AA\ (FWHM) for unbinned and binned spectra respectively, values that would be obtained for a point source to allow for discrete emission in good seeing conditions.
The upper limit was given by the observation that gas in the narrow line region has a maximum velocity width of 2,000 km~s$^{-1}$ \citep{crenshaw_connection_2005}. We locked the [N~II] $\lambda\lambda$6584/6548 line ratio to 2.95, as per \cite{osterbrock_astrophysics_2006}. The velocity widths and centroids of the lines in each component were fixed to the same value given that they arise from the same knot of gas. The integrated flux for each component must be at least 3$\sigma$ greater than the continuum noise per spectral resolution element. 

BEAT determines the best fit model in two stages: First, BEAT calculates the best fit model for a variable number of Gaussians (i.e. it finds the best 0, 1, 2, and 3 component fits); second, BEAT determines the number of components resulting in the best overall fit. In the first stage, BEAT iterates through the parameter space and calculates a likelihood for each model.  The likelihoods from all models are then summed to give a Bayesian evidence value \textit{\textbf{Z}}; the likeliest model is plotted along with the ln(\textit{\textbf{Z}}) value (see Figure~\ref{fig:beat}). The data are fit with an increasing number of Gaussians, with BEAT determining a best fit model for each. In the second stage, BEAT compares the models with different numbers of components using the Bayesian likelihood criteria: 

\begin{equation}\label{evidence}
\mathrm{ln}\left[\frac{Z_{i+1}}{Z_{i}}\right] \ge 5
\end{equation}
where $i$ corresponds to the number of components in the model. If this criterion is met, the model with $i+1$ components is 99.3\% more likely \citep{feroz_importance_2019} than the model with $i$ components, and the more complex model is adopted as the best fit. If this criterion is not met, then the simpler model with fewer components is selected. Note that reduced-$\chi^2$ fitting is not used to find the best fit between models, as a smaller $\chi^2$ value may be due to either the presence of another line component or overfitting  \citep{fischer_gemini_2017}. Using the Bayesian likelihood as the best-fit metric minimizes overfitting, resulting in a more statistically robust fit.

For the best model at each position along the slit, BEAT returns the wavelength centroid, line width, and peak flux of each component. For each of these parameters, BEAT also returns the error in the form of the standard deviation of the distribution of all generated models. These values are used to calculate the radial velocity, FWHM, and integrated flux of each ionized gas cloud at each position. These attributes (which are discussed in \S \ref{sec:kin}) are crucial to understanding the gas kinematics.

\subsection{Spectral Fitting for Other Emission Lines} \label{sec: other spectral fitting}

BEAT is designed to fit only a few emission lines at a time to obtain the kinematic components at each position. To fit the additional needed emission lines, we used the H$\alpha$ fits determined by BEAT as templates. This procedure is described in detail in \cite{revalski_quantifying_2018, revalski_quantifying_2018_2, revalski_quantifying_2021, meena_radiative_2021}, and aims to ensure that we fit the same kinematic components in each emission line for a given slit position. 
We kept the relative velocity centroid and width the same for each component, and allowed the heights of the lines to vary (locking the lines with fixed atomic ratios) to obtain the emission-line fluxes. 
We applied an SNR cutoff of integrated flux 3 times the continuum's standard deviation to ensure strong fits (using continuum regions selected near each emission line).  We use this routine to fit the lines of H$\beta$~$\lambda$4861, [O~III]~$\lambda \lambda$4959, 5007, [O~I]~$\lambda \lambda$6300, 6364, and [S~II]~$\lambda \lambda$6716, 6731 \AA.
We did not attempt to constrain the relative intensities of the [S~II] lines, because we were only interested in determining the flux of the blend for all kinematic components, resulting in reasonably low uncertainties as shown in Section 6.

\subsection{Penalized Pixel Fitting (pPXF)}\label{sec:ppxf}

To characterize the stellar kinematics in NGC~7319, we used the Ca~II $\lambda$8542.0 \AA, the Na~I $\lambda\lambda$5889.9, 5895.9 \AA, and the Mg~I~b $\lambda\lambda\lambda$5167.3, 5172.6, 5183.6 \AA\ absorption features in the KOSMOS spectra. These features were chosen due to their prominence in most galaxy spectra \citep{dressler_studying_1984, bottema_stellar_1987}. We used the central line from the Ca~II triplet, as it was the only line strong enough to determine a significant fit. To increase the signal to noise ratio (SNR) of the spectra, we chose to bin every four rows of the 2D spectrum (projecting to 1\farcs0) along the slit.

\begin{figure*}[t]
    \centering
    \includegraphics[width=\linewidth]{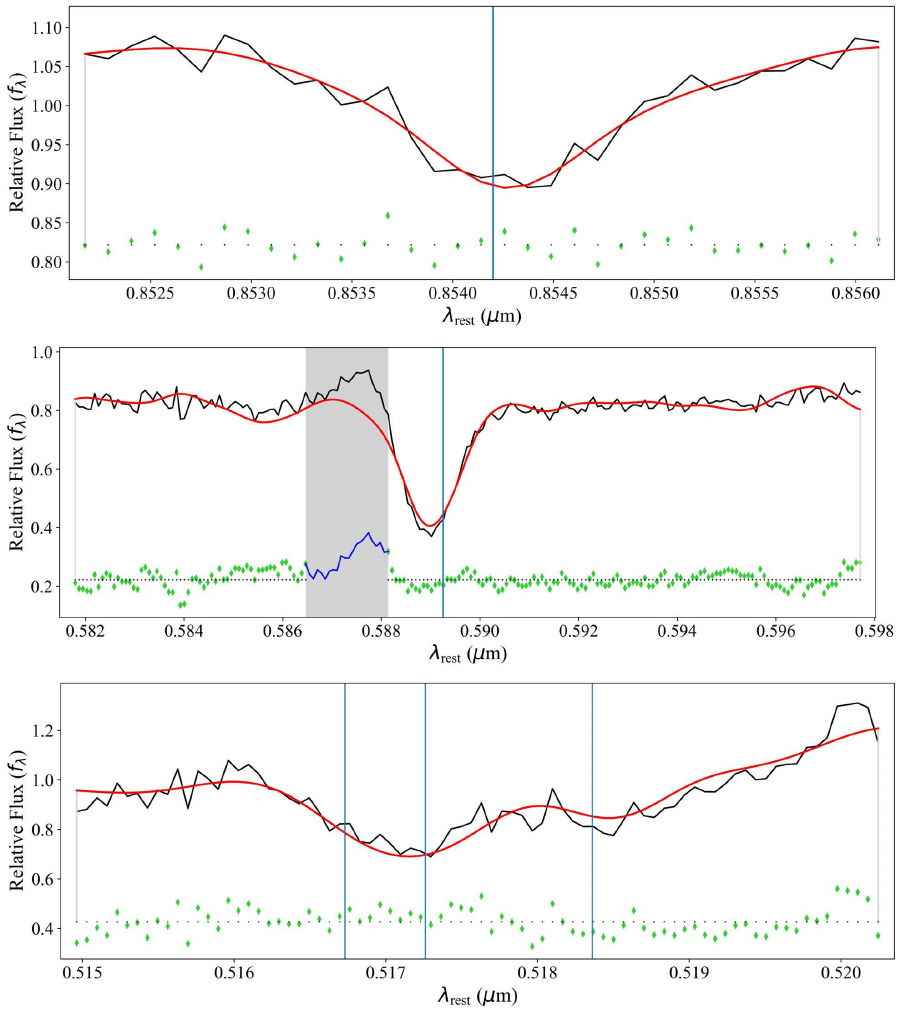}
    \caption{Example pPXF fits for NGC~7319 at a position of 1\farcs032 from the nucleus. Black is the log-rebinned spectrum. The blue vertical line is the rest wavelength of the line. Red is the initial best fit to the spectrum using the stellar templates. Green points are the residuals. \textit{Top:} Ca~II $\lambda$8542.0 \AA. \textit{Middle:} Na~I $\lambda\lambda$5889.9, 5895.9 \AA\ (blended). The masked region (grey) is the He I $\lambda$5876 \AA\ emission line. \textit{Bottom:} Mg~I~b $\lambda\lambda$5167.3, 5172.6 \AA\ (blended), and $\lambda$5183.6 \AA.}
   \label{fig:ppxf_fits}
\end{figure*}

To fit the absorption features, we used the Penalized Pixel Fitting function (pPXF; \citealp{cappellari_full_2023, cappellari_improving_2017, cappellari_parametric_2004}). pPXF determines the line-of-sight velocity distributions (LOSVD) for each 1D spectrum across the galaxy. For our template spectra, we observed the stars HD 97438 (spectral type F0 III), HD 117876 (spectral type G8 III), and HD 125560 (spectral type K3 III), as we expect the bulge to be primarily composed of G and K giants \citep{binney_galactic_1998}.
We observed these stars with the same grisms as the galaxy spectra to ensure the templates have the same wavelength coverage and dispersion as the galaxy spectra (see Table~\ref{table:obs}). By using stars we observed with the same instrumentation configuration, we avoid any of the issues that may arise from using generic spectral libraries. This method has been successfully employed before, i.e. \cite{bentz_low-mass_2016}.

We limited the wavelength range to $8520-8565$ \AA\ for the Ca~II line, to $5820-5980$ \AA\ for the Na~I lines, and to $5150-5225$ \AA\ for the Mg~I~b lines. We used two moments for the fit (representing the velocity and velocity dispersion; the spectral resolution is not high enough to merit the use of Hermitian components) and a degree of 2. Examples of fits for the Ca~II, Na~I, and Mg~I~b lines are shown in Figure~\ref{fig:ppxf_fits}. The fitting process determined the best-fit parameters (velocity and velocity dispersion) using the three stellar templates. Using this best fit as a template, we then ran a 10-iteration bootstrap \citep{davidson_wild_2008} on each spectrum, taking the mean LOSVD and the standard deviation of the 10 iterations to determine the velocity and velocity error at that position.


\section{Kinematics} \label{sec:kin}
\subsection{Stellar Kinematics}\label{sec:rotation}

\begin{figure*}
    \centering
    \includegraphics[width=0.9\textwidth]{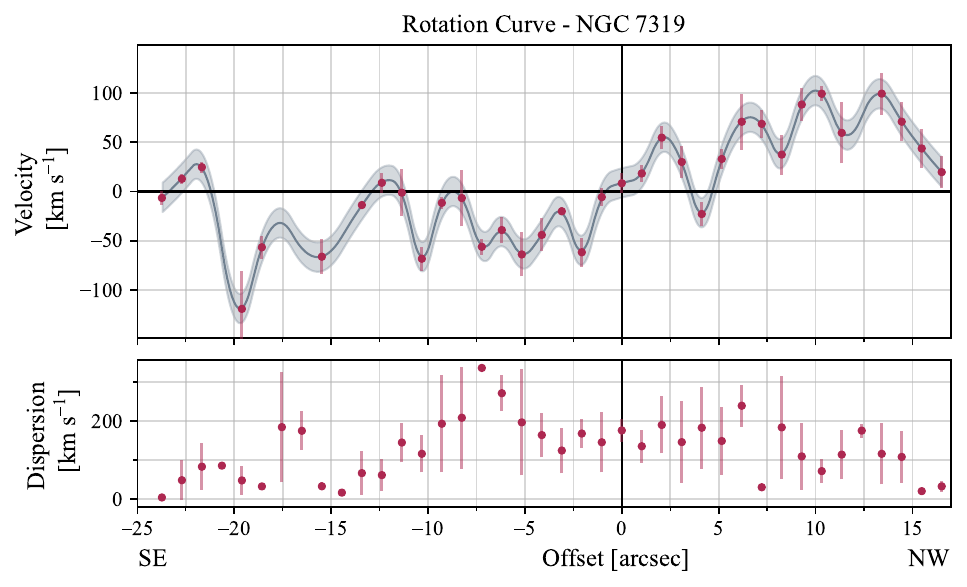}
    \caption{\textit{Top:} Rotation Curve of NGC~7319 along its major axis. The red points are the averaged velocities from the Ca~II, Na~I, and Mg~I b fits, with propagated errors. The grey curve is an interpolated fit. The shaded area of the curve is the average error of $\pm$15~km~s$^{-1}$. \textit{Bottom:} Velocity Dispersions, averaged across the three lines.} 
   \label{fig:rotation}
\end{figure*}

To determine the stellar rotation curve of NGC~7319, we averaged the Ca~II, Na~I, and Mg~I fits from \S \ref{sec:ppxf}, and propagated the errors accordingly. To smooth and interpolate the points, we used an univariate second order spline. For the error on the smoothed rotation curve, we chose to use the average of the errors on all the points, approximately 15~km~s$^{-1}$. The resulting rotation curve for NGC~7319 is shown in Figure~\ref{fig:rotation}. While the rotation curve is much more variable than a typical one, the kinematics of NGC~7319 are known to be disturbed. \cite{yttergren_gas_2021} shows a similarly bumpy rotation field. To the northwest, we see primarily low-velocity redshifted values, with two peaks at 2\arcsec\ and 7\farcs5 of $\sim$50~km~s$^{-1}$. To the southeast, we see similarly low-velocity blueshifted values, with a peak approaching 0~km~s$^{-1}$ at 3\arcsec. These features are all reflected in the rotation curve created by \cite{yttergren_gas_2021}. 

\subsection{Ionized Gas Kinematics}

\begin{figure*}
    \centering
    
    \subfloat[][]{\includegraphics[width=0.48\textwidth]{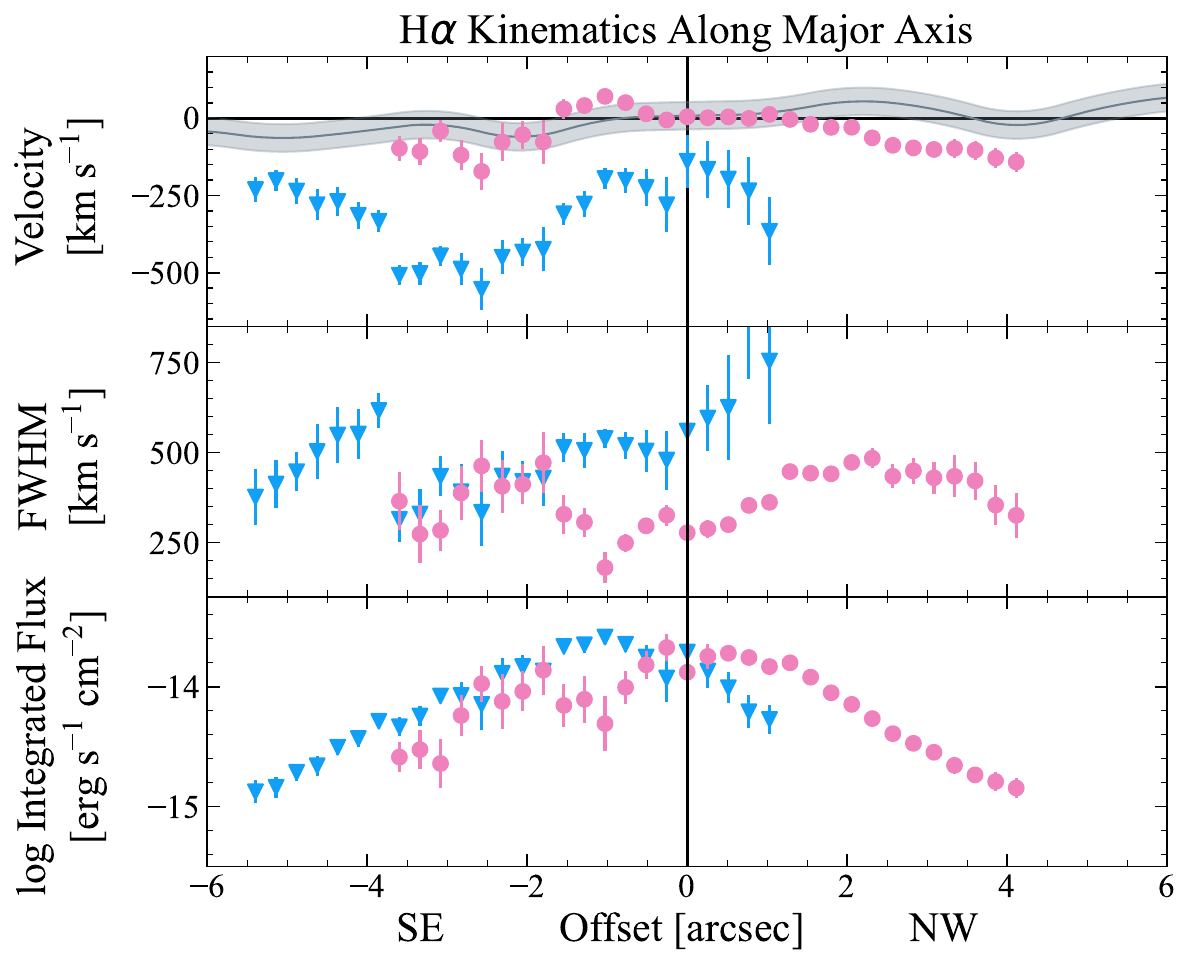}}
    \hfill
    \subfloat[][]{\includegraphics[width=0.48\textwidth]{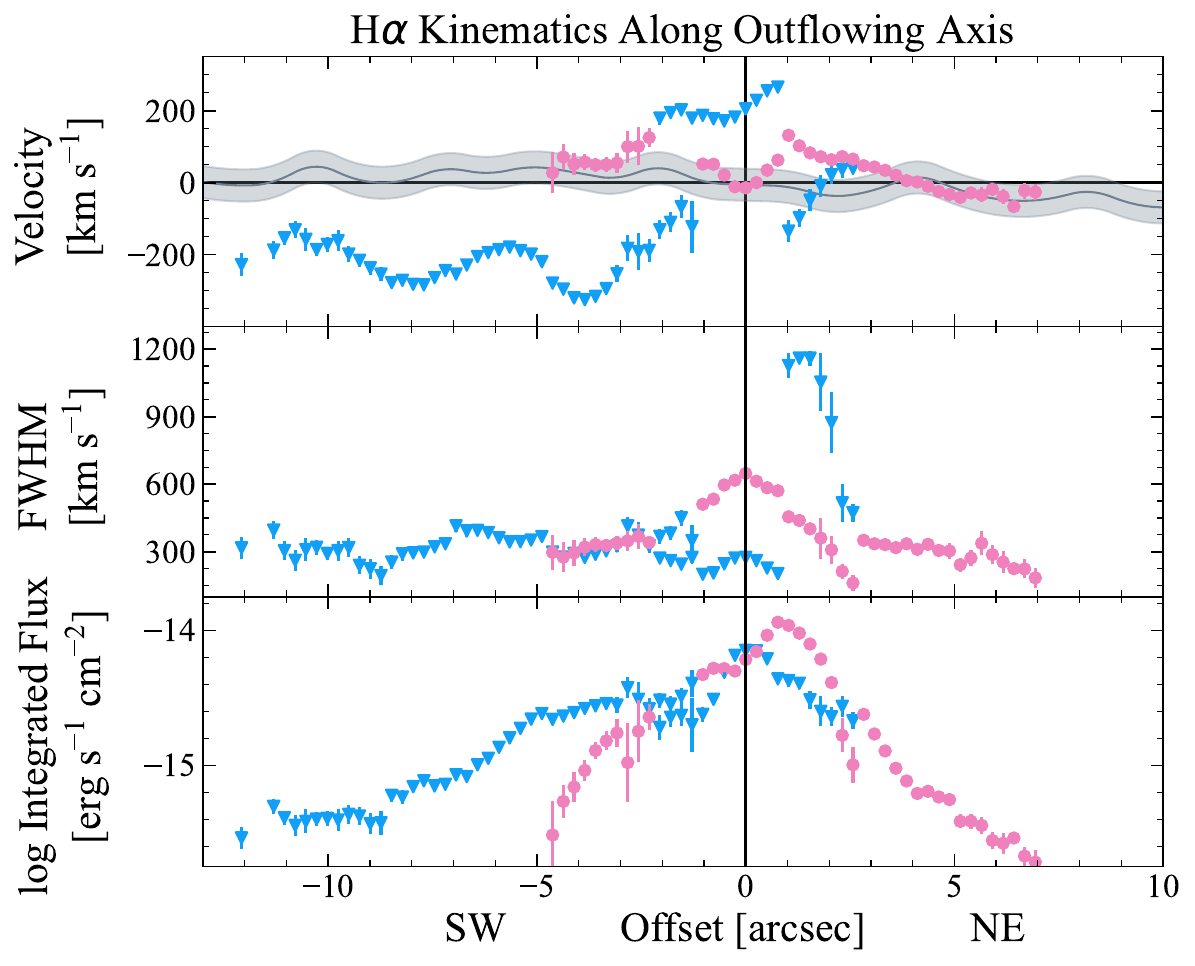}}\\
    \subfloat[][]{\includegraphics[width=0.48\textwidth]{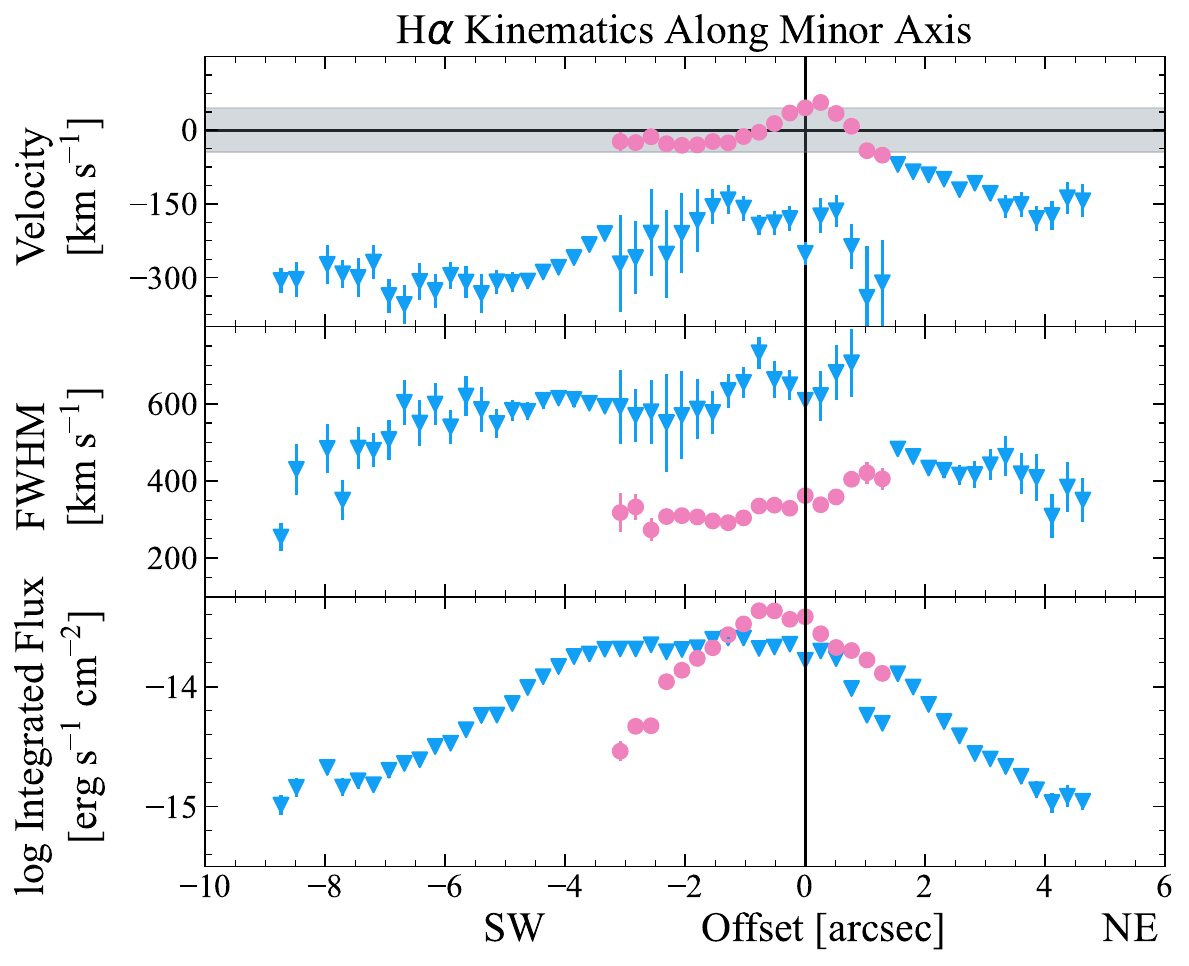}} 
    \hfill
    \subfloat[][]{\includegraphics[width=0.48\textwidth]{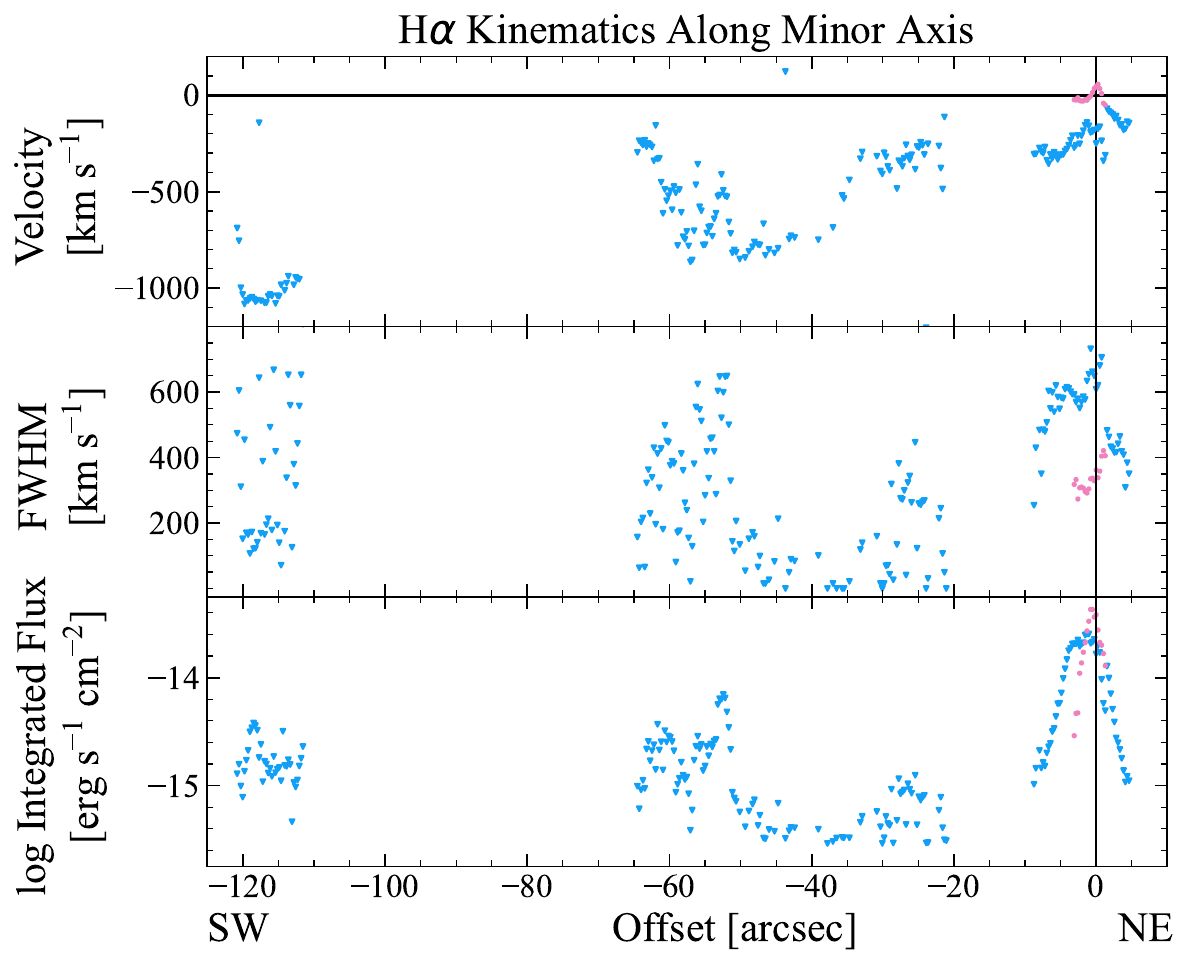}} \\ 
    
    \caption{H$\alpha$ kinematic plots along the \textit{(a)} major axis of NGC~7319, \textit{(b)} outflowing axis of the NLR,  \textit{(c)} minor axis of NGC~7319 in the nuclear region, and \textit{(d)} extended region. In each figure, the top panel shows the radial velocity distribution of H$\alpha$ emission, the middle panel shows the FWHM distribution of each component, and the bottom panel shows the integrated flux distribution. The kinematic components are sorted into rotational motion in pink (circles), and non-rotational motion in blue (triangles). The grey curve running through all velocity plots is the projected rotation curve of NGC~7319.}
    
    \label{fig:kinematics}
\end{figure*}

Beat has the ability to fit the broadline region (BLR) -- for NGC 7319, BEAT determined that a 0 component fit was the best fit for the BLR, so we treat NGC 7319 as a Seyfert 2. For the narrow lines, BEAT returned a maximum of two components for each fit, although fits of up to three components were allowed.
Figure~\ref{fig:kinematics} shows the ionized gas kinematics for H$\alpha$ along our three KOSMOS slits, in the reference frame of NGC~7319. In each plot, the top panel shows the radial velocity distribution of the H$\alpha$ emission, the middle panel shows the FWHM distribution of each component, and the bottom panel shows the integrated flux distribution.   The two-component fits are seen near the nucleus, where the kinematics are more complex. Farther from the nucleus, we see only one-component fits. The grey curve in each plot is the rotation curve of NGC~7319, as described in \S \ref{sec:rotation}. 
The maximum and minimum velocity values, and the overall shape of the rotation curve, vary according to the projection of the rotation curve at each position angle. 

To separate the likely rotational from non-rotational motion in the emission lines, we used two criteria. For points to be classified as rotation, they must fall near the rotation curve (within $\approx$100 km~s$^{-1}$) and have a low FWHM ($\lesssim$400~km~$^{-1}$). For points that did not meet but were close to these criteria, we classified them by kinematic groupings; i.e. we looked for discontinuities in velocity, FWHM, or flux to distinguish between groups. Points we classified as primarily rotation are shown as pink circles in Figure~\ref{fig:kinematics}. The blue triangular points are more difficult to classify. Possibilities include AGN-powered outflows, gravitationally-powered tidal flows, or some combination thereof. 

Along the major axis, there is unlikely to be significant tidal influence, as the points are perpendicular to the bridge of gas. In this case, we see likely outflows up to $-$550 km~s$^{-1}$ at a distance of 5\farcs4 (2.6 kpc) to the southeast. This is consistent with the result from \cite{aoki_high_1996}, who found a maximum blueshifted outflow radial velocity of $-$500 km~s$^{-1}$. The mean blueshifted velocity we find is 320 $\pm$ 23 km~s$^{-1}$. Along the minor axis to the northeast, the points are opposite the tidal flow and also unlikely to show tidal influence. In this region, the outflows reach a peak radial velocity of $-$180~km~s$^{-1}$ and an extent of 4\farcs5 (2.2 kpc). To the southwest in both the minor and outflowing axes, we cannot separate out the AGN outflows from the tidal flows using the kinematic plots alone. Distinguishing between the two is discussed more in \S \ref{sec:disc}.
Finally, there are significant redshifted velocities up to $+$250~km~s$^{-1}$ close to and on either side of the nucleus, which help to provide further constraints on kinematic models of the outflows.

\begin{figure*}
    \centering
    \includegraphics[width=0.9\textwidth]{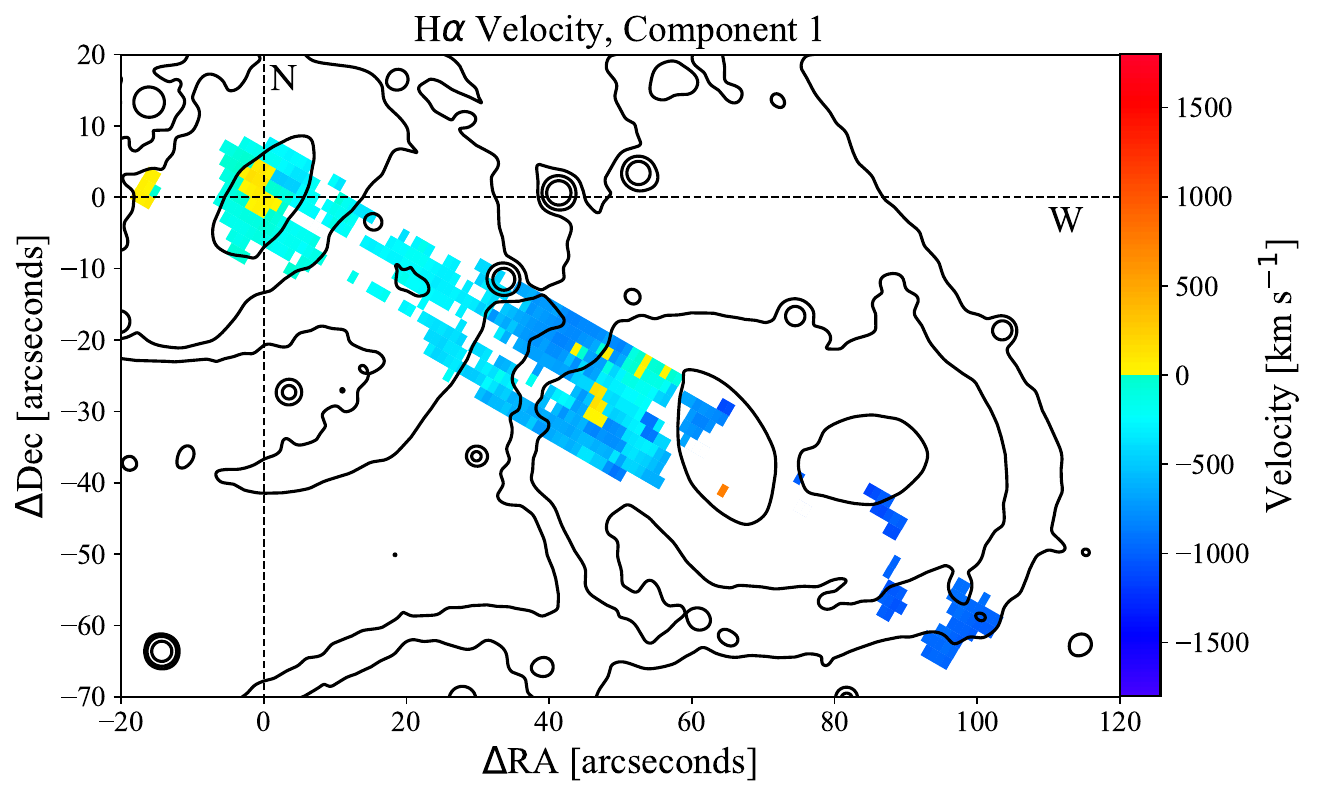}
    \includegraphics[width=0.9\textwidth]{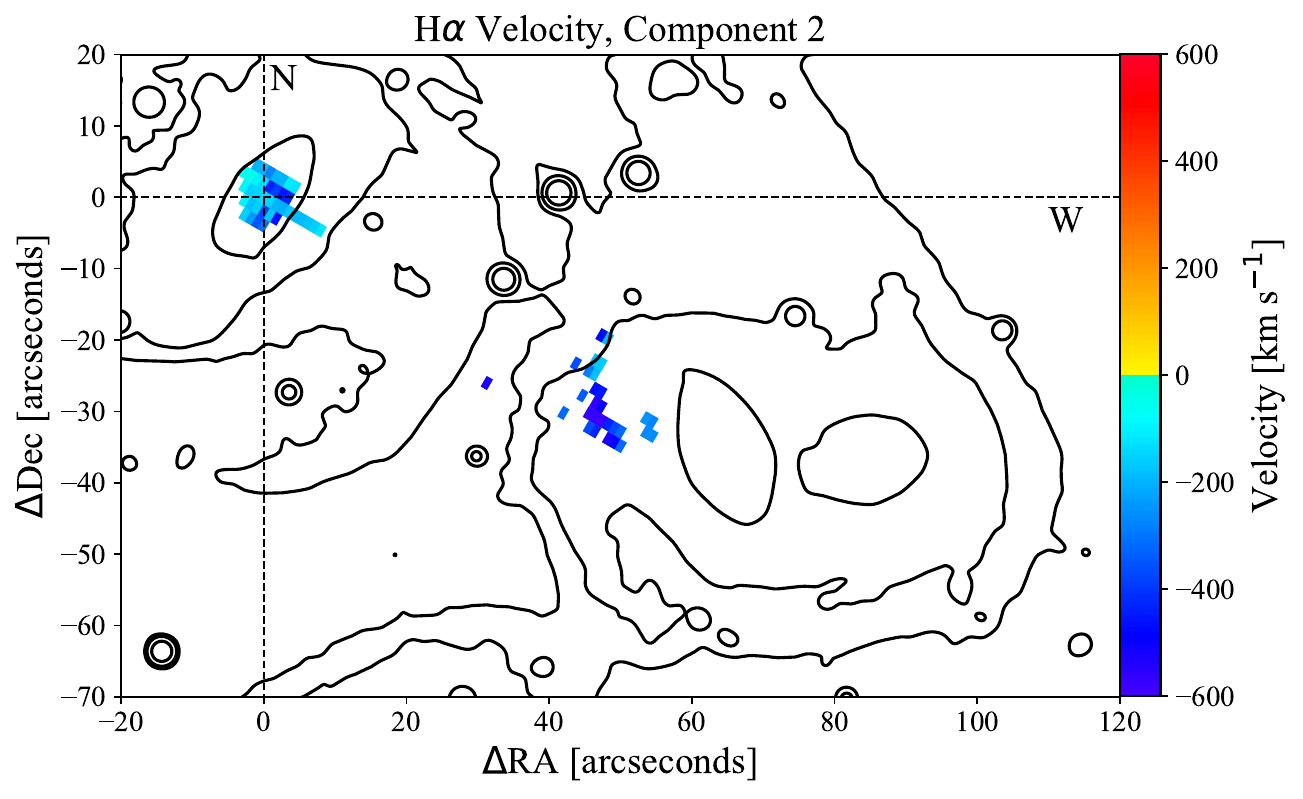}
    \caption{2D velocity maps of the H$\alpha$ kinematics in Stephan's Quintet. The background contours (black) were created using the HST F814W filter with contour levels of (0.05, 0.07, 0.2) electrons s$^{-1}$, which were chosen to bring out key features of the group. The slits are centered along the minor axis of NGC~7319, with points along the slit colored by velocity (given relative to the rest frame of NGC~7319).   The top map represents the first component fits only, as sorted by width (lowest widths; FWHM $= 70-850$~km~s$^{-1}$), and the bottom map represents the second component fits (FWHM $= 130-1130$~km~s$^{-1}$). Points along the slits that are not colored had no significant fits returned by BEAT (or only 1 component fits, in the case of the bottom plot), indicating weak or no ionized gas at the position.}
    \label{fig:2D_kinematic}
\end{figure*}

To help us further analyze the AGN outflows and tidal flows, we created a 2D kinematic map of the velocities by including the offset slits along the minor axis (shown in Figure~\ref{fig:slits}). Figure~\ref{fig:2D_kinematic} shows the kinematic maps, one for each component (sorted by width, with the lower widths as the first component). To increase the SNR, particularly along the bridge, we binned the data by four spectral rows along the cross-dispersion direction to obtain a good balance between spatial resolution (1\farcs0) while increasing the SNR by a factor of $\sim$2. The plots show the RA offset and Dec offset in arcseconds, centered on the nucleus of NGC~7319. We used HST continuum images to create contours of the galaxies for orientation (black lines). The points along the slit are colored by velocity, in the reference frame of NGC~7319. From these maps, we can see there is blueshifted and low-amplitude redshifted ionized gas in the nuclear region of NGC~7319 (consistent with the 1D kinematic plot along the minor axis in Figure~\ref{fig:kinematics}b). The bridge of gas is also evident, with high blueshifted velocities reaching up to $\sim$950 km~s$^{-1}$. There also appears to be two regions of ionized gas in the spiral arms of NGC~7319 ($\sim$$-$15\arcsec\ northeast) and NGC~7318B ($\sim$100\arcsec\ southwest). To truly disentangle the outflows from tidal flows, we need to create a geometric model of the NLR. 

\section{Biconical Model} \label{sec:bicone}

Biconical-outflow models have been successfully used to empirically describe the NLR kinematics of many nearby Seyfert galaxies \citep{antonucci_spectropolarimetry_1985, pogge_extended_1988, pedlar_radio_1993, schmitt_anisotropic_1994, nelson_space_2000, fischer_determining_2013, meena_investigating_2023}.
To determine the best fit bicone model for NGC~7319, we began by generating models using the kinematic modeling code developed by \cite{das_mapping_2005}. This code creates models based on input parameters including the bicone's orientation on the sky, inclination with respect to the plane of the sky, minimum and maximum half-opening angle (HOA), and height along the axis. Additionally, the models incorporate a velocity profile that begins at 0~km~s$^{-1}$ in the center and increases linearly to a turnover radius, after which the velocity decreases linearly to zero at a given maximum height along the bicone axis. Many Seyfert NLRs have been shown to match this empirical velocity law quite well \citep{fischer_determining_2013}.

\begin{figure*}
    \centering
    \subfloat{\includegraphics[width=0.8\textwidth]{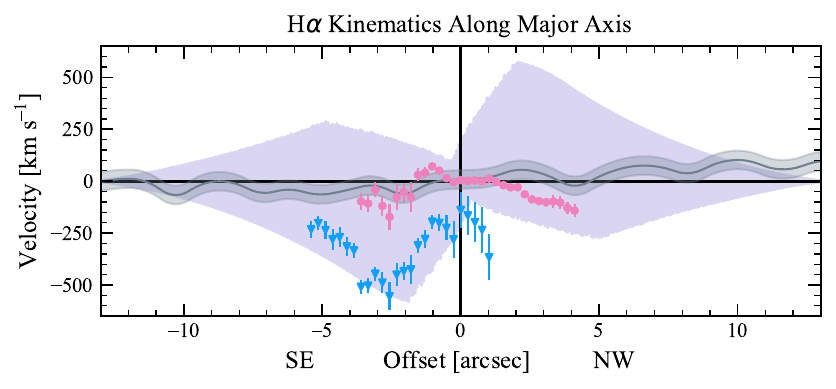}} \\
    \subfloat{\includegraphics[width=0.8\textwidth]{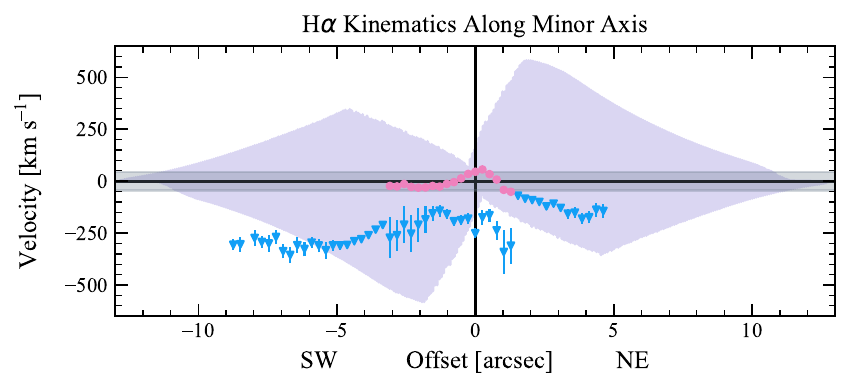}} \\
    \subfloat{\includegraphics[width=0.8\textwidth]{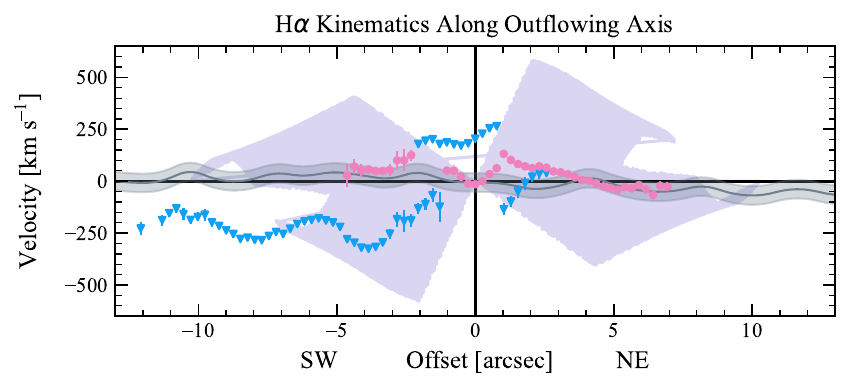}}
    
    \caption{H$\alpha$ kinematic plots with symbols and color scheme as in Figure~\ref{fig:kinematics}. The ``0" position represents the nucleus as defined by the continuum peak. The purple shaded regions show the model velocity areas from the bicone model, and represent the range of velocities produced by the bicone model at each position.}
    \label{fig:bicone-kinem}
\end{figure*}

We began by determining a range of values for each input parameter from the H$\alpha$ image (shown in Figure~\ref{fig:slits}) and the kinematic plots (shown in Figure~\ref{fig:kinematics}) of the nuclear region. 
Initial values for the outer opening and position angles and extent of the bicone were determined from the geometry of the H$\alpha$ image. Initial values for the turnover radius, maximum velocity, and inclination of the bicone were determined from the kinematic plots. When choosing our paremeters, we also took into account the major axis of galaxy of 148\degree\ \citep{jarrett_2mass_2003}, as well as the 56\degree\ SW inclination of galaxy \citep{jarrett_2mass_2003, pereira-santaella_low-power_2022}.
Our initial parameter space is shown in Table~\ref{table:bicone space}, resulting in a total of over 200,000 models. We implemented conditions that eliminate some of these models, such as the turnover radius must be less than the height, and our line of sight must be outside of the bicone to give a Seyfert 2 view. We allowed our models to extend beyond the edges of the major and minor axis slits; nevertheless, the code determined the maximum possible HOA to be very near to these axes. For each of these models, the \cite{das_mapping_2005} code produces a range in possible velocities as a function of projected location from the AGN, due to the range in opening angles that provide the thickness of the bicone. The bicone kinematics are extracted from the model at each slit position, and plotted against the kinematic points from BEAT. This ``model velocity space'' is shown by the purple shaded envelopes in Figure~\ref{fig:bicone-kinem}.

\begin{figure*}
     \centering
     \includegraphics[width=0.9\linewidth]{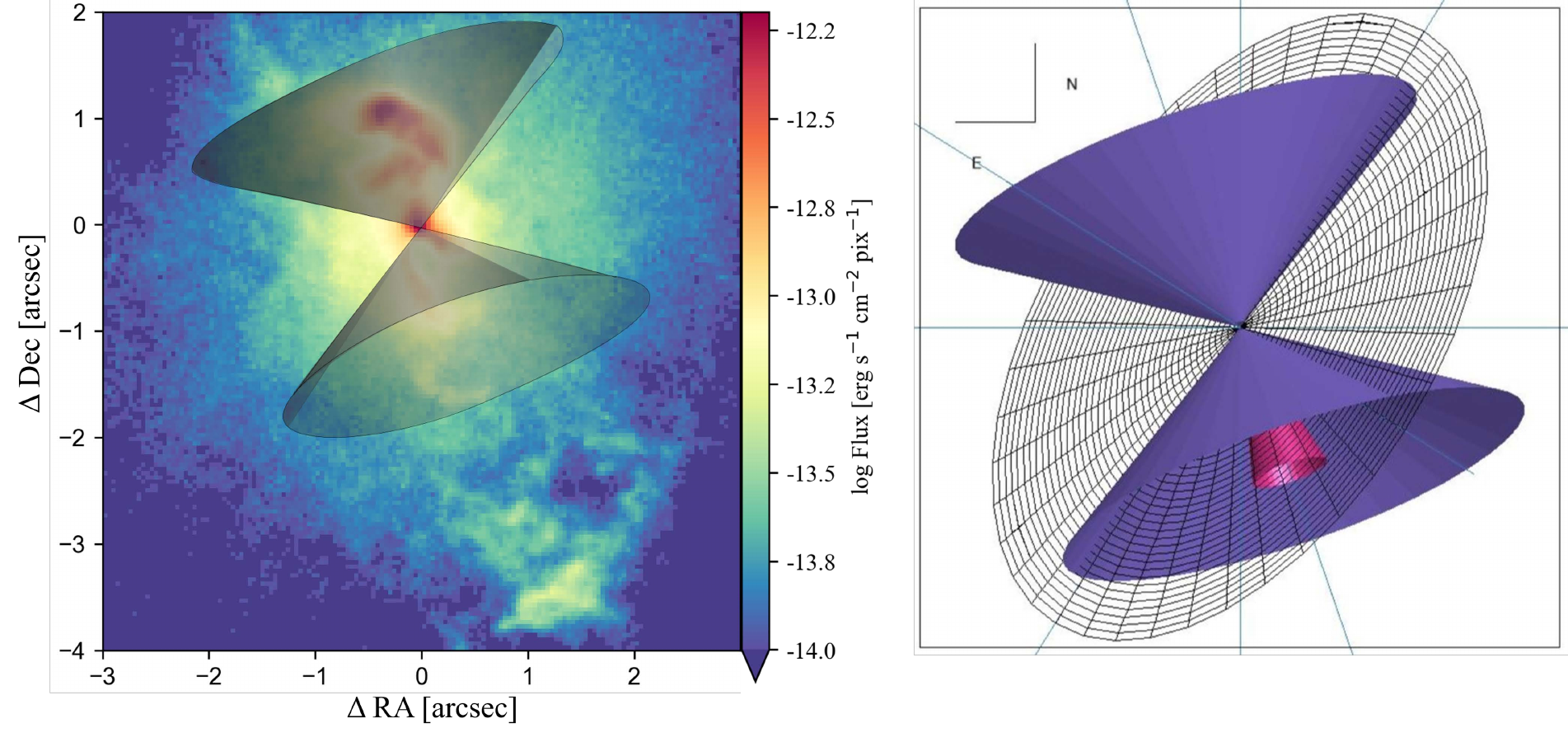}
     \caption{Bicone model of NGC~7319. \textit{Left:} The bicone model plotted over an H$\alpha$ image (same as in Figure~\ref{fig:slits}).The darker shaded regions are in front of the disk of the galaxy, while the lighter shaded regions are obscured by the galactic disk. \textit{Right:} A 3D rendering of the bicone model. The grid represents the disk of the galaxy, the purple cone the outer HOA (55\arcdeg), and the pink cone the inner HOA (14\arcdeg).}
     \label{fig:bicone}
\end{figure*}

From these kinematic models, we chose the optimal model using the algorithm described in \cite{falcone_analysis_2024}. The algorithm calculates the fraction of data points that fall within the model velocity space and characterizes how well the model fits the data. Because our data exhibits both rotation and outflows, we did not include any points that we deemed due to rotation in this calculation (pink circular points in Figure~\ref{fig:bicone-kinem}). We did, however, include all the rest of the kinematic points (blue triangular points) in the calculation, despite the possibility that some of these likely have tidal influences in addition to the AGN influence. The algorithm determines which model has the most kinematic points that fall within the model velocity space. The model velocity areas are weighted to prevent the widest opening angle from always being chosen, as it has the largest model area and would therefore capture the most points. After determining the best model, we generated a new set of models over a smaller, finer parameter space and ran the algorithm again. The final parameter models are listed in Table~\ref{table:bicone params}, along with the host galaxy parameters and associated references.
The final bicone model is shown in Figure~\ref{fig:bicone}, overlaid over the H$\alpha$ image. The darker shaded regions of the bicone are in front of the disk of the galaxy, while the lighter shaded regions are obscured by the galactic disk.

The southwest portion of the bicone is pointed towards our line of sight. From the shaded regions in Figure~\ref{fig:bicone-kinem}, we see that the bicone model predicts high-amplitude blueshifted velocities and low-amplitude redshifted velocities in the south. We see kinematic points within the predicted blueshift velocities in the south quadrant for all three position angles. However, the corresponding redshift velocity south quadrants have few to no kinematic points. In Figure~\ref{fig:bicone}, we see that this quadrant is obscured by the galactic disk, so the absence of points in the optical is to be expected $-$ follow-up observations in the infrared may reveal these points. Similarly, the bicone predicts an antisymmetric pattern in the north, predicting low-amplitude blueshifted velocities and high-amplitude redshifted velocities. Once again, we see the predicted blueshifted points, but the redshifted space is sparse, as it is also obscured by the galactic disk.
There are some redshifted points close to and on either side of the nucleus consistent with the biconical outflow model, which may be visible due to their apparent brightness and/or low extinction near the nucleus. 
There are some kinematic points that fall outside the model entirely, particularly to southwest along the minor and outflowing axes. These deviations from the model are discussed more in \S\ref{sec:disc}.

\begin{table}
\centering
\hspace{-1.5cm}   
\begin{tabular}{|r|c|c|} \hline
    Parameters & Range & Interval \\
    \hline\hline

    Bicone PA [\arcdeg] & 17 -- 23 & 1 \\    
    Bicone Inclination [\arcdeg] & 12 -- 17 & 1\\
    Inner HOA [\arcdeg] & 12 -- 22 & 1 \\
    Outer HOA [\arcdeg] & 53 -- 58 & 1\\
    Turnover Radius [kpc] & 2.4 -- 3.4 & 0.2 \\
    Max Height [kpc] & 2.8 -- 4.0 & 0.2 \\
    Max Velocity [km s$^{-1}$] & 600 -- 750 & 50 \\
  \hline
\end{tabular}
\caption{Initial parameter space of the bicone models. Rows are (1) major axis of bicone axis, (2) inclination of bicone axis relative to plane of sky, (3) minimum HOA, (4) maximum HOA, (5) turnover radius, (6) maximum height of bicone along its axis, and (7) maximum outflow velocity, located at the turnover radius.}
\label{table:bicone space}

\end{table}

\begin{table}
\centering
\hspace{-1cm}   
\begin{tabular}{|r|l|} \hline
  Parameters & Value  \\ 
  \hline
  \hline
  
  Galaxy PA &  $148$\arcdeg \\
  Galaxy Inclination & $56$\arcdeg (SW closer) \\ [5pt]
  Bicone PA &  $19_{-1}^{+2}$\arcdeg \\ [5pt]
  Bicone Inclination &  $14_{-1}^{+3}$\arcdeg  \\ [5pt]
  Inner HOA & $16_{-2}^{+6}$\arcdeg \\ [5pt]
  Outer HOA & $55_{-1}^{+1}$\arcdeg \\ [5pt]
  Turnover radius & $3.10_{-0.66}^{+0.24}$ kpc \\ [5pt]
  Max height & $3.52_{-0.08}^{+0.24}$ kpc \\ [5pt]
  Max velocity &  $700_{-70}^{+15}$ km s$^{-1}$ \\ [5pt]
  
  \hline
\end{tabular}

\caption{Final Bicone parameters for NGC 7319. Rows are (1) major axis of galaxy from \cite{jarrett_2mass_2003}, (2) inclination of galaxy from \cite{jarrett_2mass_2003}, direction of inclination from \cite{pereira-santaella_low-power_2022}, (3) major axis of bicone axis, (4) inclination of bicone axis relative to plane of sky, (5) minimum HOA, (6) maximum HOA, (7) turnover radius, (8) maximum height of bicone along its axis, and (9) maximum outflow velocity, located at the turnover radius. Errors for this work are reported as a 95\% confidence interval.}

\label{table:bicone params}
\end{table} 

\section{Line Ratio Analysis} \label{sec:BPT}
\begin{figure*}
    \centering
    \includegraphics[width=\textwidth]{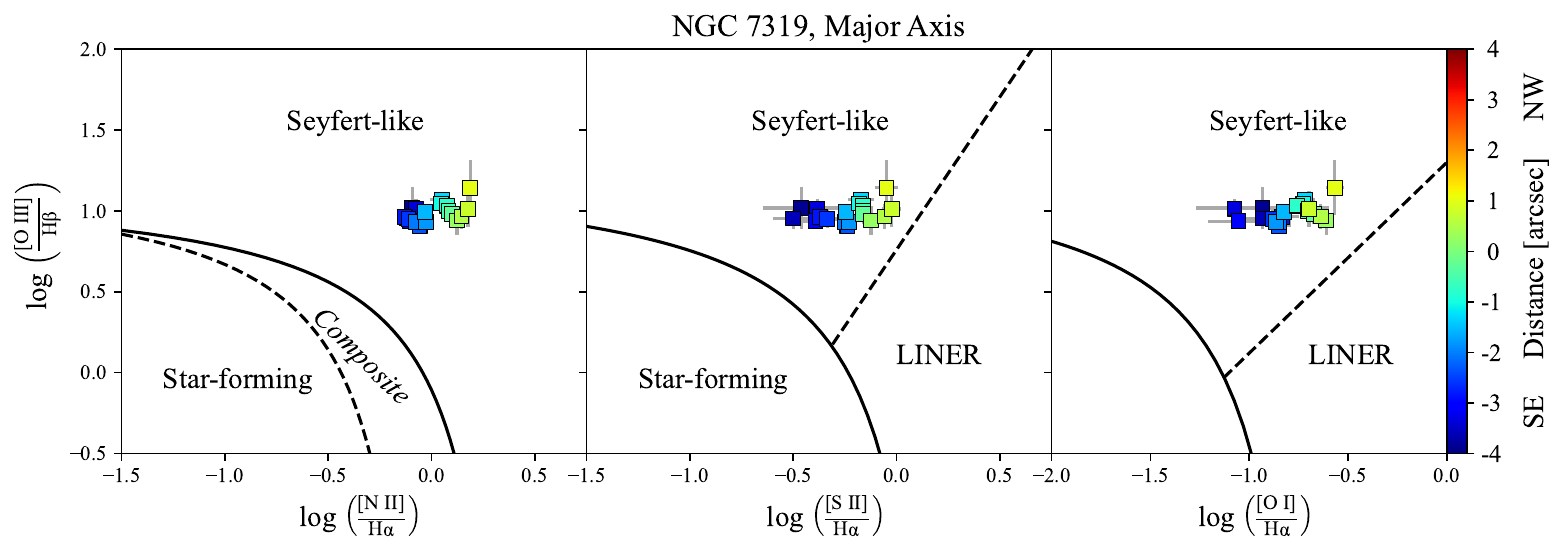}\\
    \medskip
    \includegraphics[width=\textwidth]{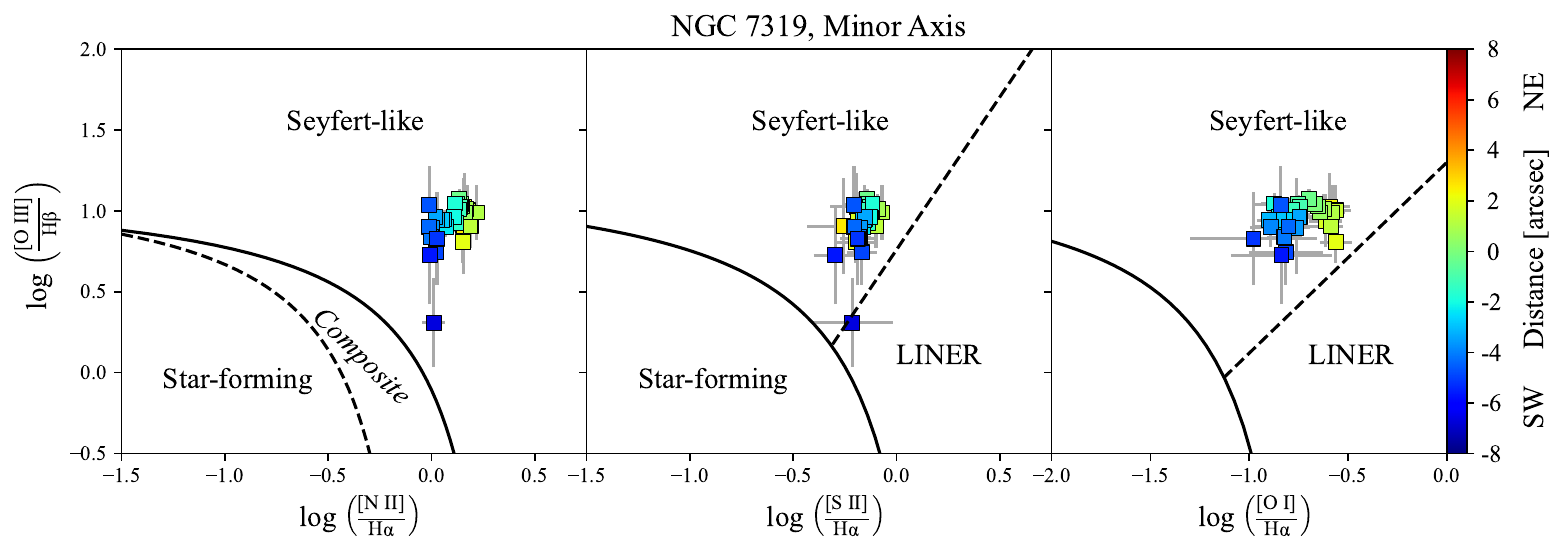}\\
    \medskip
    \includegraphics[width=\textwidth]{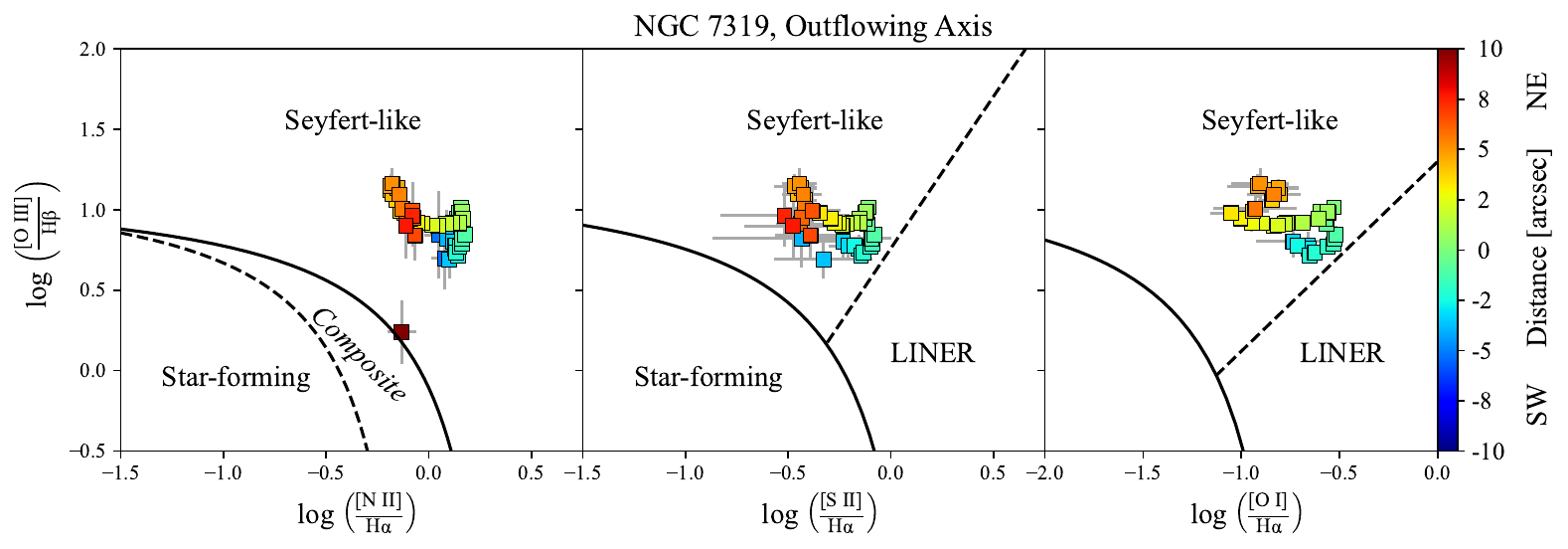}\\
    \medskip
    
    \caption{BPT diagrams for the nuclear region of NGC~7319, along the major (\textit{top}) and minor (\textit{middle}) axes of NGC~7319, and the outflowing axis of the NLR (\textit{bottom}). For positions that have multiple components, the fluxes from the components have been summed together. Symbols are colored according to their distances from the central SMBH.}
    \label{fig:BPT}
\end{figure*}

We used emission line measurements to create Baldwin-Phillips-Terlevich (BPT) diagrams \citep{baldwin_classification_1981, veilleux_spectral_1987} for NGC~7319 and the IGrM. A BPT diagram is an important diagnostic tool that involves comparing line ratios of [O~III]/H$\beta$ to [N~II]/H$\alpha$, [O~I]/H$\alpha$, and [S~II]/H$\alpha$ to distinguish whether gas at a particular distance from the nucleus is ionized by the AGN, by star formation, or both. This analysis allows us to trace the physical extent of the ionizing radiation as a function of distance from the nucleus, and quantify the influence of AGN feedback processes within Stephan's Quintet.
We note that the demarcation lines between different ionizing sources in this type of analysis are not always sharp \citep{ji_constraining_2020} and that there may be transition regions where both sources of photoionization are important.
Furthermore, BPT analysis does not account for shock ionization unless additional parameter(s) are used, as discussed below. Thus, we use this analysis to get a general sense of the dominant ionization sources in each region.
The resulting BPT diagrams along the major axis and the outflowing axis are shown in Figure~\ref{fig:BPT}. The demarcations that separate the various classifications are described in \cite{kewley_theoretical_2001, kewley_host_2006} and \cite{kauffmann_host_2003}. 

Along the major axis (top panel of Figure~\ref{fig:BPT}), the ionizing source is firmly Seyfert-like within $+$1/$-$4\arcsec; some points from the kinematic plots in Figure~\ref{fig:kinematics} were lost due to the SNR cutoff. For the minor axis (middle panel of Figure~\ref{fig:BPT}), we limit the diagram to only the nuclear region of NGC~7319 (i.e. only points represented in Figure~\ref{fig:bicone-kinem}). We still see Seyfert-like ionization, with two points verging on low-ionization nuclear emission-line region (LINER) ionization to the southwest (towards the bridge). The extended emission along the minor axis is discussed further below. Along the outflowing axis (bottom panel of Figure~\ref{fig:BPT}), we see strong Seyfert-like ionization in the nuclear region, with the farthest points at $\sim$11\arcsec\ in the northeast verging on composite ionization in the [N~II] BPT, indicating a possible contribution from star formation.

For the minor axis, we created 2D BPT diagrams (Figure~\ref{fig:2D_BPT}). These plots have the same layout as the 2D kinematic plots in Figure~\ref{fig:2D_kinematic}, with the addition of Chandra ACIS-S X-ray contours (shaded in light grey) obtained from \dataset[Chandra ObsId 7924]{https://doi.org/10.25574/07924}. For each slit position, we created a traditional BPT diagram, then classified each point into one of the three categories. In each of the three BPT diagrams, we see predominantly Seyfert-like ionization out to a distance of $\sim$13\arcsec\ (green dashed circle). This corresponds to the maximum extent of outflowing gas we expect to see based on the bicone model. The [S~II] and [O~I] BPT diagrams show a few points as LINER and star forming ionization within this radius -- this is likely due to weaker ionizing fluxes at these distances from the AGN. We see two star forming regions at $\sim$15\arcsec\ northeast of the AGN, and at $\sim$100\arcsec\ to the southwest in a spiral arm of NGC~7318B, both of which are  observed by \cite{fedotov_star_2011} and \cite{gallagher_hubble_2001}. In the bridge region, we see a mix of ionizing sources, which are not completely consistent among the three diagnostics, indicating an ionization mechanism outside of the traditional BPT sources. The majority of the bridge is aligned with the Chandra X-ray contours, indicating that the bridge is shock ionized \citep{xu_physical_2003, osullivan_chandra_2009, appleton_multiphase_2023}. The bridge also has very high FWHM values (see Figure~\ref{fig:2D_BPT}d), which is another indicator of shock ionization \citep{zhu_theoretical_2025}.
The results are consistent with those of \cite{arnaudova_weave_2024}, who perform much more detailed analysis of the large-scale shock front and associated structures in Stephan's Quintet.

\begin{figure*}
    \centering
    
    \subfloat[][]{\includegraphics[width=0.8\textwidth]{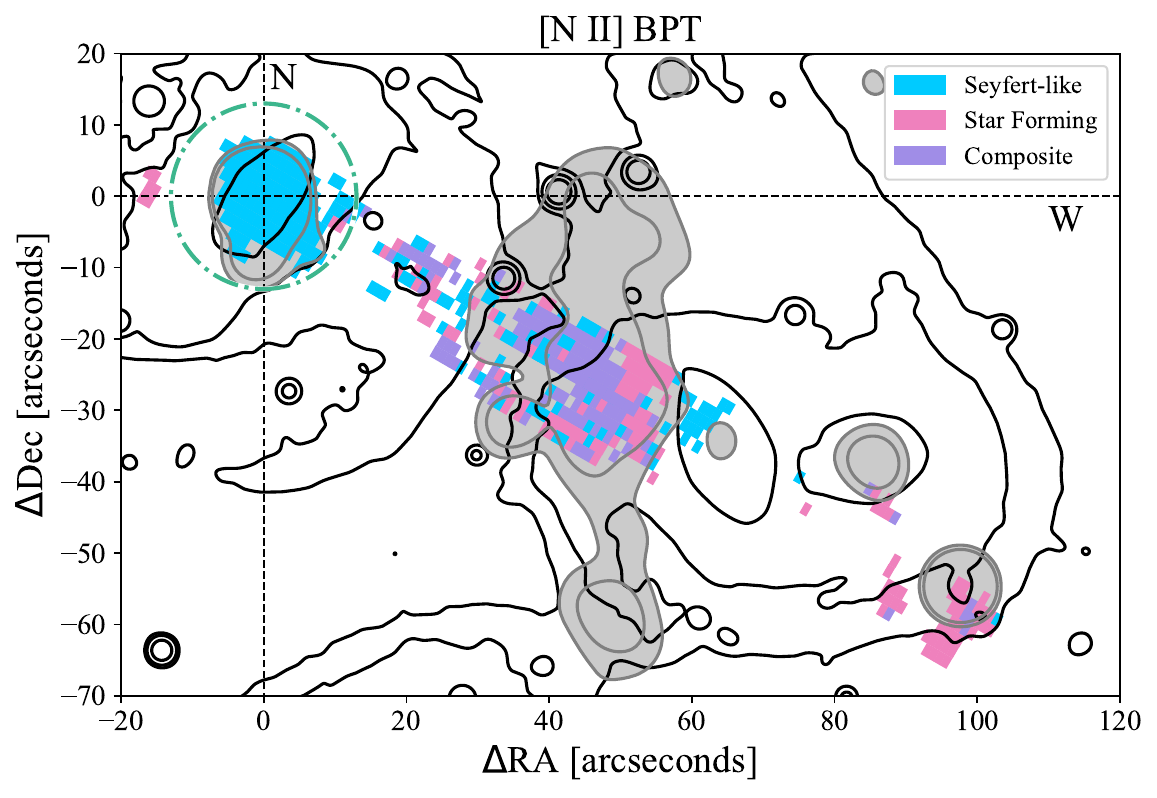}}\\
    \subfloat[][]{\includegraphics[width=0.8\textwidth]{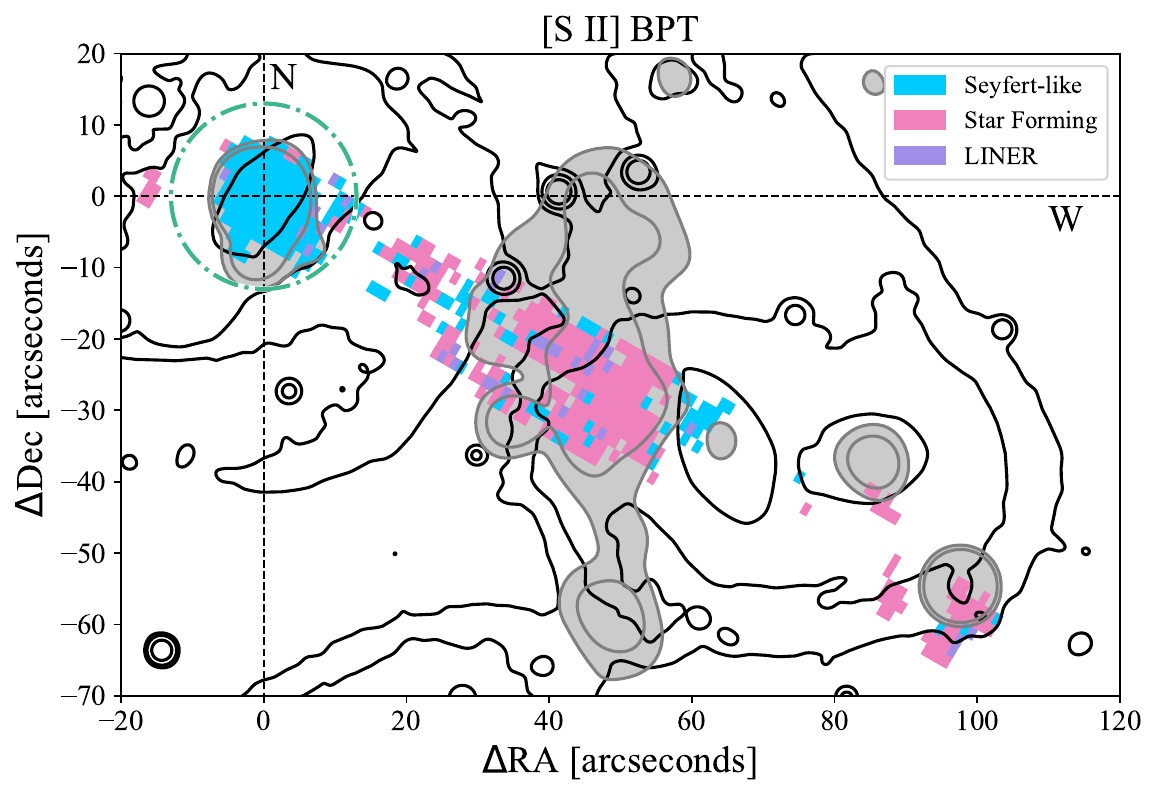}} \\ 
\end{figure*}

\begin{figure*}
    \ContinuedFloat
    \centering
    \subfloat[][]{\includegraphics[width=0.8\textwidth]{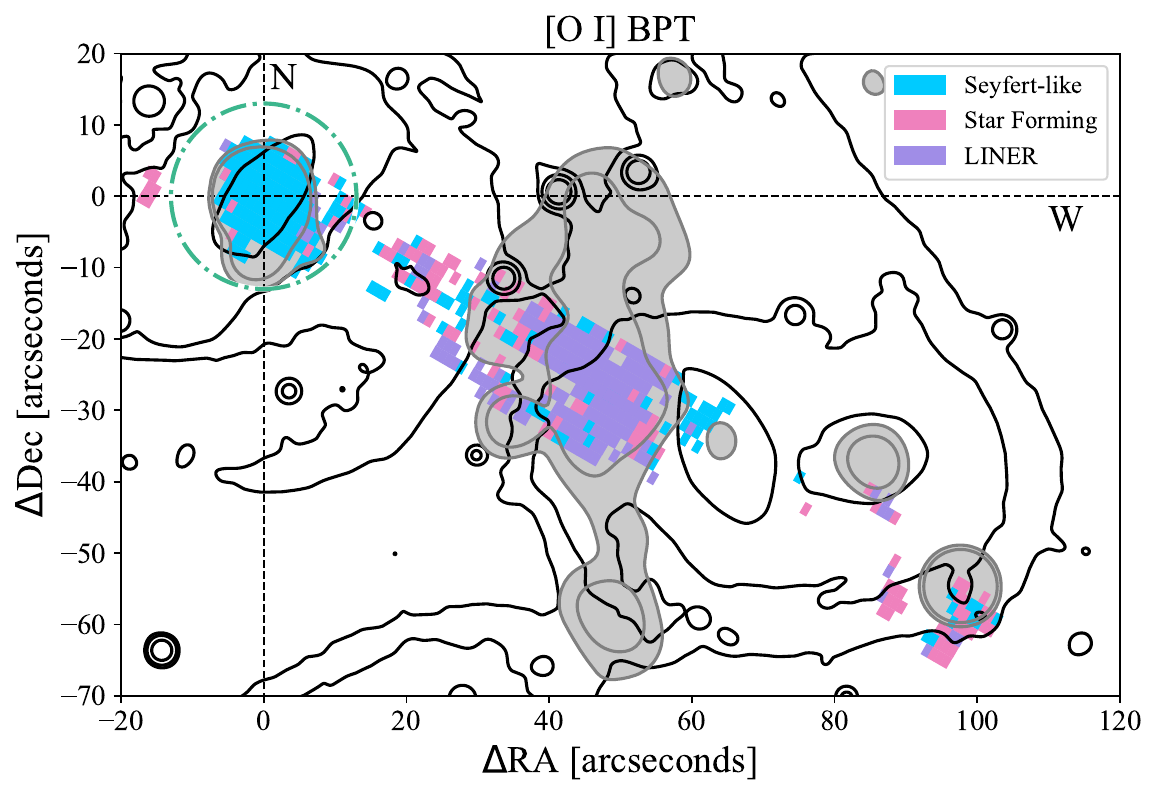}}\\
    \subfloat[][]{\includegraphics[width=0.8\textwidth]{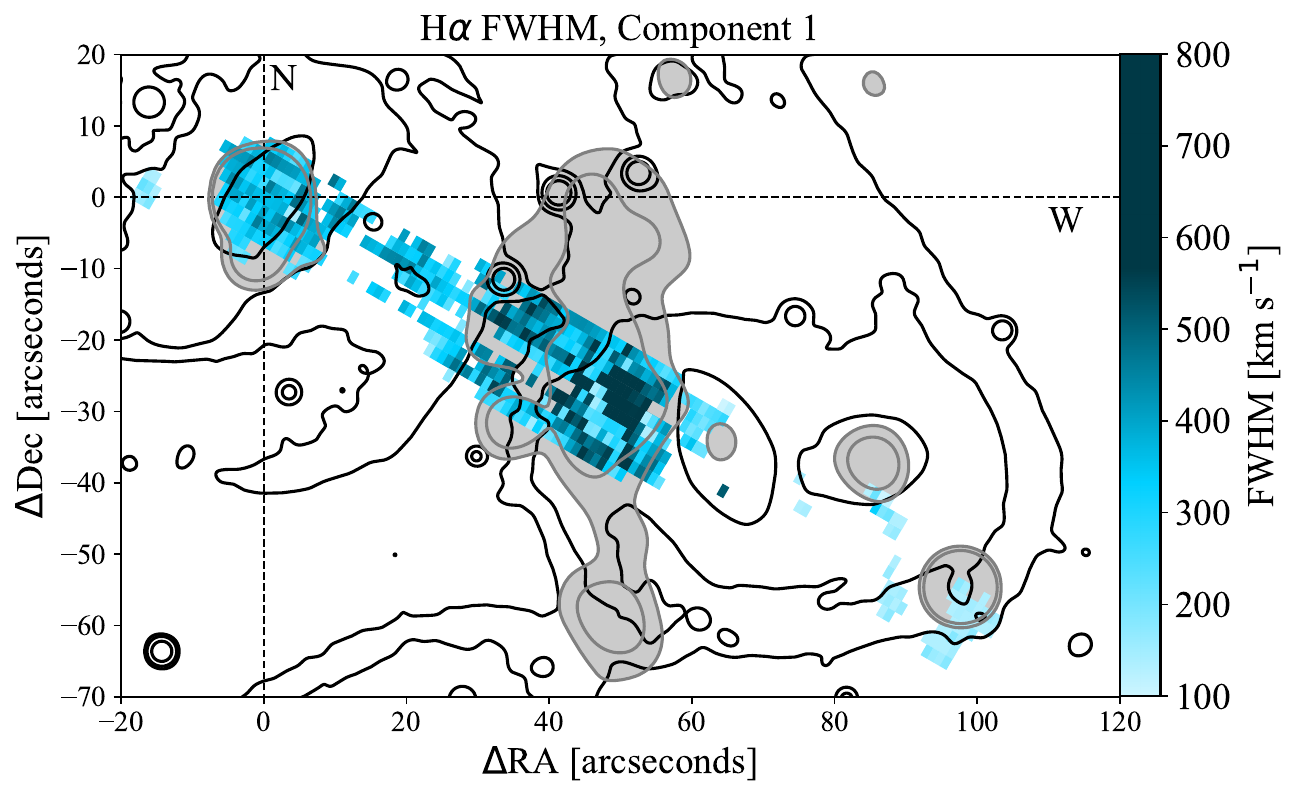}}\\
    
    \caption{2D representations of the BPT diagrams along the minor axis slits. Grey shaded regions are contours from Chandra data, indicating shock-ionized regions. The green dashed line is a 13\arcsec\ radius, the maximum extent of AGN-ionized gas we expect to see based on the bicone model. The diagrams represent \textit{(a)} [N~II] BPT, \textit{(b)} [S~II] BPT, \textit{(c)} [O~I] BPT, and \textit{(d)} FWHM. Note the high FWHM values in the shocked region, and the low FWHM in the star-forming regions at 15\arcsec\ NE and 100\arcsec\ SW.}
    \label{fig:2D_BPT}
\end{figure*}

\pagebreak
\section{Results \& Discussion} \label{sec:disc}
We have shown that the AGN in NGC~7319, which is interacting with other nearby galaxies and embedded in the intragroup medium in Stephan's Quintet, exhibits outflows of ionized gas similar to those seen in Seyfert galaxies outside of compact groups \citep{revalski_quantifying_2022, meena_investigating_2023, falcone_analysis_2024}. The similarities are striking, in that all of these AGN show biconical outflows close to the nucleus with increasing velocity to a turnover point, followed by a decrease to systemic velocity on scales of hundreds to thousands of parsecs. 
Here we consider the transition between AGN outflows and tidal flows in a compact group in more detail, and investigate additional similarities (or differences) with AGN that are not in compact group, particularly as a function of luminosity. 

\subsection{Outflow -- Tidal Flow Transition}

Based on the observed kinematics of the emission lines in NGC~7319 and the matching biconical outflow model, we see AGN-driven gas out to a projected distance of $\sim$13\arcsec\ (6.3~kpc), where the velocity model returns to 0~km~s$^{-1}$ in Figure~\ref{fig:bicone-kinem}. Our independent emission-line ratio analysis seen in Figure~\ref{fig:2D_BPT} shows that the entire nuclear region shows Seyfert-like ionization out to this same projected distance of $\sim$13\arcsec. Along the major axis, which is perpendicular to the bridge (see Figure~\ref{fig:slits}), we do not find any additional kinematic points that would indicate tidal influence. In this case, our kinematic points stay within the bicone model velocity space, and closely follow the turnover point (top panel of Figure~\ref{fig:bicone-kinem}). The outflows only reach a maximum extent of 5\farcs4 in this direction. However, along the minor and outflowing axes, which intersect with the bridge of gas, our kinematic points show deviations from the bicone models.  Along the minor axis (central panel of Figure~\ref{fig:bicone-kinem}), we do not observe a turnover, but rather a flattened velocity curve, with kinematic points falling outside the model velocity space beginning around 5\arcsec\ to the southwest (in the direction of the bridge).  Furthermore, if we look at the FWHM in  Figure~\ref{fig:kinematics}c, we can see a break, which indicates a disturbance. To the northeast, on the opposite side of the nucleus from the bridge, we would not expect any tidal influence, and the outflows extend to 4\farcs6. Along the outflowing axis (bottom panel of Figure~\ref{fig:bicone-kinem}), we see similar deviations from the bicone. Kinematic points fall outside the bicone model velocity space beginning at 7\arcsec; however, we see unexpected velocities beginning at 5\arcsec. The kinematic points follow the expected turnover, but at 5\arcsec\ there is an additional acceleration instead of a return to zero velocity, which is inconsistent with the biconical outflow model. Again, if we return to Figure~\ref{fig:kinematics}b, we can see a clear break in the FWHM at 5\arcsec\, which marks a kinematic disturbance.

Taking all these factors together, we surmise that while the ionized gas in NGC~7319 is both AGN-ionized and AGN-driven out to 13\arcsec\ (6.3 kpc), there are only undisturbed outflows out to $\sim$5\arcsec\ (2.4 kpc), at which point there is an additional tidal influence driving the motions of the gas.
These results are consistent with other studies that show that outflows of AGN-ionized gas affect only the central regions of their host galaxies at low to moderate luminosities \citep{holden_no_2025, revalski_quantifying_2025}, and in this case, do not directly affect the galaxy group as a whole.

Although we have identified the regions of interaction between the AGN outflows and tidal flows in NGC~7319, the exact roles of the tidal flows in feeding the AGN and the AGN outflows in potentially arresting the tidal flows require further investigation.
Additional kinematic studies at higher angular resolution (i.e., ($\sim$0\farcs1) of multiple gas phases (molecular, atomic, ionized) are likely required.
To these ends, detailed analyses of the existing MIRI IFU observations of NGC~7319 and Stephan's Quintet will be helpful.
\subsection{Turnover Correlation with AGN Luminosity}

Although NGC~7319 is a Seyfert 2 galaxy and its central region cannot be observed directly, the average bolometric luminosity over the NLR can be calculated from the relation $L_{\mathrm{bol}}=3500 \times L_{5007}$ \citep{heckman_present-day_2004}. We found the total [O~III] luminosity across the outflowing axis (the PA capturing the majority of the [O~III] emission) to be $L_{5007}~=$~6.2~$\times~
10^{41}$~erg~s${^{-1}}$, corresponding to $L_{\mathrm{bol}}~=~2.2~\pm~0.3~\times~10^{45}$~erg~s${^{-1}}$. This is about an order of magnitude brighter than the value listed by \cite{woo_active_2002} of $L_{\mathrm{bol}}~=~1.55~\times~10^{44}$~erg~s${^{-1}}$.

\begin{figure}[h]
    \centering
    \includegraphics[width=\linewidth]{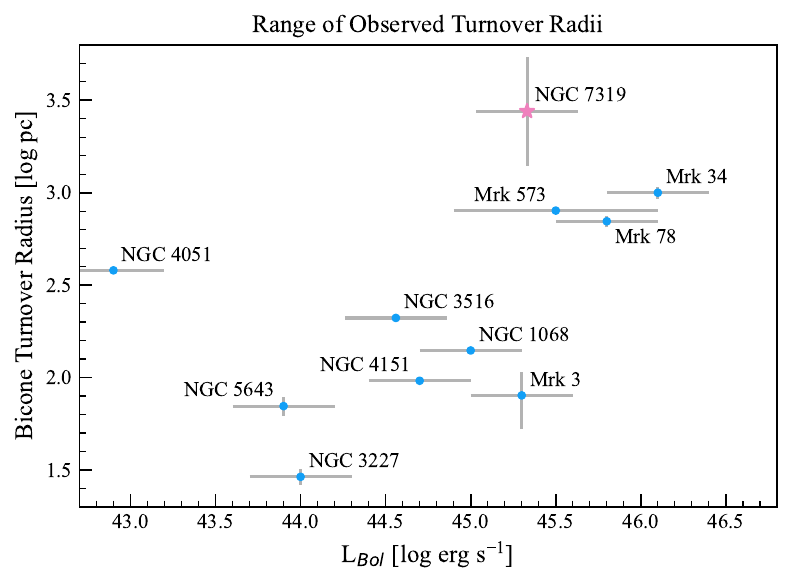}
    \caption{A comparison of observed turnover radius and bolometric luminosity. Adapted from Figure 9 of \cite{meena_investigating_2023}, with additional points of NGC~3227 from \cite{falcone_analysis_2024} and NGC~3516 from \cite{tutterow_shape_2025}. Error bars on the turnover radii are based on the resolution of the HST Space Telescope Imaging Spectrograph, or  KOSMOS in the case of NGC~7319. The uncertainties in the bolometric luminosities are $\pm$0.3 dex for all targets except Mrk 573, which is $\pm$0.6 dex (see \citealp{meena_investigating_2023}). While NGC~7319 has a relatively high turnover radius for its luminosity, it still supports the conclusion by \cite{meena_investigating_2023} that more luminous AGN show higher turnover radii.}
    \label{fig:turnover_lum}
\end{figure}

We can compare the combination of bicone turnover radius and bolometric luminosity to other nearby AGN. Figure~\ref{fig:turnover_lum}, adapted from \cite{meena_investigating_2023}, shows the positive correlation between $L_{\mathrm{bol}}$ and bicone turnover radius. The majority of points are from \cite{meena_investigating_2023}, with additional points of NGC~3227 from \cite{falcone_analysis_2024} and NGC~3516 from \cite{tutterow_shape_2025}. NGC~7319 has a somewhat higher turnover radius than AGN with similar bolometric luminosities, but it still follows the trend of increasing turnover radius with higher AGN luminosity. This trend can be naturally explained by radiative driving of the NLR gas \citep{meena_investigating_2023, falcone_analysis_2024, tutterow_shape_2025}. 

\pagebreak
We calculated the SMBH mass using the following equation from \cite{tremaine_slope_2002}:

\begin{equation}
    log \left( \frac{M_{BH}}{M_\odot}\right) = 4.02~log\left( \frac{\sigma}{200} \right) + 8.19
\end{equation}
Using the velocity dispersion from the inner 5\arcsec, we find a black hole mass of $M_{BH} = 6.5 \pm 3.6 \times 10^7 M_\odot$, which is slightly greater than the mass of $M_{BH} = 2.40 \times 10^7 M_\odot$ obtained by \cite{woo_active_2002}.

Our black hole mass and bolometric luminosity yield an Eddington ratio of 0.26.
Interestingly, this relatively high ratio based on the average luminosity over the extent of the NLR ($\sim$25,000 light years) implies that NGC~7319 has been accreting at a relatively high rate over at least $\sim$25,000 years, on average, possibly as a consequence of the massive tidal flows of gas in Stephan's Quintet.

\section{Conclusions} \label{sec:conc}
We obtained long-slit optical spectra of the Seyfert 2 AGN NGC~7319 and the IGrM of Stephan's Quintet using the APO KOSMOS. We created 
galactic rotation models, biconical outflow models, and kinematic and ionization maps of the nuclear and bridge region with the aim of disentangling rotation, AGN outflows, and tidal flows.  Our main conclusions are as follows: 

\begin{enumerate}

    \item We generated a rotation curve of NGC~7319 based on stellar absorption features that extends out to a projected distance of 15\arcsec\ (7.2 kpc). Within the inner 7\arcsec\ (3.4~kpc) of the central SMBH, we were able to separate two kinematic components of ionized gas: galactic rotation and AGN-driven outflows. 

    \item We developed the first biconical outflow model of the ionized gas in the NLR of NGC~7319. The bicone has an inclination of 14\degree\ relative to the plane of sky, with an inner HOA of 16\degree\ and an outer HOA of 55\degree. These values result in much of the far side of the bicone being obscured by the galactic disk, explaining the paucity of redshifted gas velocities seen in the optical. The bicone model has a velocity turnover radius of $\sim$3.1 kpc, and extends to a distance of $\sim$6.3 kpc along the bicone. NGC~7319 follows the general trend of increasing turnover radius with increasing AGN luminosity.
    
    \item We observed undisturbed AGN-powered outflows out to  a projected distance of 5\farcs4 (2.6 kpc) that reach a maximum velocity of 550~km~s$^{-1}$.  
    We found that between 5\arcsec~--~13\arcsec\ (2.4~--~6.3~kpc), the ionized gas motion is likely powered by a combination of AGN outflows and tidal flows. Past 13\arcsec, the gas motion is most likely powered by tidal flows, which extend along the bridge of ionized leading to NGC~7318B, the current interloper in Stephan's Quintet.
    
    \item We performed a line ratio analysis and determined the gas shows Seyfert-like ionization out to a projected distance of 13\arcsec\ (6.3 kpc), which matches the region of AGN-driven outflows from the kinematic model. 
       
\end{enumerate}

 Although the extent of the AGN outflows and the regions of AGN interaction with the tidal flows in Stephan's Quintet are now well constrained, further study is needed to explore the nature of these interactions and their connection to AGN feeding and feedback in compact groups of galaxies.

\clearpage
\begin{acknowledgments}

We thank the anonymous reviewer for their helpful comments on this paper.

Many of the observations used in this paper were obtained with the Apache Point Observatory 3.5-meter telescope, which is owned and operated by the Astrophysical Research Consortium.

Some of the data presented in this work are based on observations with the NASA/ESA Hubble Space Telescope and were obtained from the Mikulski Archive for Space Telescopes (MAST), which is operated by the Association of Universities for Research in Astronomy, Incorporated, under NASA contract NAS5-26555. These observations are associated with program number \href{https://archive.stsci.edu/proposal_search.php?mission=hst&id=11502}{11502}. The specific observations used in this work can be accessed via DOI: \dataset[10.17909/78sx-6q77]{http://dx.doi.org/10.17909/78sx-6q77}. HST observations were also obtained from the Hubble Legacy Archive, which is a collaboration between the Space Telescope Science Institute (STScI/NASA), the Space Telescope European Coordinating Facility (ST-ECF/ESA) and the Canadian Astronomy Data Centre (CADC/NRC/CSA).

The scientific results reported in this article are based in part on data obtained from the Chandra Data Archive (program number 7924). The specific observations used in this work can be accessed via DOI: \dataset[10.25574/07924]{https://doi.org/10.25574/07924}. 

This research has made use of NASA’s Astrophysics Data System. IRAF is distributed by the National Optical Astronomy Observatories, which are operated by the Association of Universities for Research in Astronomy, Inc., under cooperative agreement with the National Science Foundation.

This research has made use of the NASA/IPAC Extragalactic Database \dataset[(NED)]{https://catcopy.ipac.caltech.edu/dois/doi.php?id=10.26132/NED1}, which is operated by the Jet Propulsion Laboratory, California Institute of Technology, under contract with the National Aeronautics and Space Administration.
\end{acknowledgments}

\bibliography{references}{}
\bibliographystyle{aasjournal}

\end{document}